\def\spose#1{\hbox to 0pt{#1\hss}}

\def\multleft#1{\hbox to size{\vbox {\halign {\lft{##}\cr #1}}\hfill}\par}
\def\multright#1{\hbox to size{\vbox {\halign {\rt{##}\cr #1}}\hfill}\par}

\def\today{\ifcase\month\or January\or February\or March\or April\or May\or
      June\or July\or August\or September\or October\or November\or December\fi
      \space\number\day, \number\year}
\def\s{\hbox{\phantom{5}}}	

\def\cm{{\rm\thinspace cm}}

\def\erg{{\rm\thinspace erg}}

\def\K{{\rm\thinspace K}}
\def\keV{{\rm\thinspace keV}}
\def\km{{\rm\thinspace km}}
\def\kpc{{\rm\thinspace kpc}}
\def\Lsun{\hbox{$\rm\thinspace L_{\odot}$}}

\def\Mpc{{\rm\thinspace Mpc}}
\def\Msun{\hbox{$\rm\thinspace M_{\odot}$}}
\def\pc{{\rm\thinspace pc}}

\def\s{{\rm\thinspace s}}
\def\ks{{\rm\thinspace ks}}
\def\yr{{\rm\thinspace yr}}

\def\ergpcmsqps{\hbox{$\erg\cm^{-2}\s^{-1}\,$}}
\def\emSB{\hbox{$\erg\cm^{-2}\s^{-1}\rm{arcsec}^{-2}\,$}}

\def\ergps{\hbox{$\erg\s^{-1}\,$}}

\def\kmps{\hbox{$\km\s^{-1}\,$}}

\def\Msunpyr{\hbox{$\Msun\yr^{-1}\,$}}
\def\pcm{\hbox{$\cm^{-3}\,$}}

\def\kmpspMpc{\hbox{$\kmps\Mpc^{-1}$}}

\def\H2{\hbox{H$_{2}$}}

\documentclass[usegraphicx]{mn2e}

\usepackage{times}
\usepackage{amssymb}
\include{defn}

\begin{document}
\hsize=6truein

\title[Integral field spectroscopy of H$\alpha$ emission in cooling flow cluster cores]{Integral field spectroscopy of H$\alpha$ emission in cooling flow cluster cores: disturbing the molecular gas reservoir
 \thanks{Based on observations performed at the European Southern Observatory, Chile (Programme ID: 71.A-3014(A) and at the William Herschel Telescope, La Palma (Programme ID: SW2005b01). $\dagger$ Email: r.j.wilman@durham.ac.uk}}
\author[R.J.~Wilman, A.C.~Edge and A.M.~Swinbank]
{\parbox[]{6.in} {R.J.~Wilman$^{\dagger}$, A.C.~Edge and A.M.~Swinbank\\ \\
\footnotesize
Department of Physics, University of Durham, South Road, Durham, DH1 3LE. \\ }}
\maketitle

\begin{abstract}
We present optical integral field spectroscopy of the H$\alpha$-luminous ($>10^{42}$\ergps) central cluster galaxies in the cores of the 
cooling flow clusters A1664, A1835, A2204 and Zw8193. From the [NII]+H$\alpha$ complex in these moderate resolution (70--150\kmps) spectra we 
derive two-dimensional views of the distribution and kinematics of the emission line gas, and further diagnostics from the [SII] and [OI] lines. 

The H$\alpha$ emission shows a variety of disturbed morphologies, ranging from smooth but distorted to clumpy and filamentary, with 
velocity gradients and splittings of several hundred \kmps~on spatial scales of 20\kpc~or more. Despite the small sample size, there are 
some generic features. The most disturbed H$\alpha$ emission appears to be associated with secondary galaxies within 10--20\kpc~(projected) of 
the central galaxy and close in velocity to the H$\alpha$. The global H$\alpha$ kinematics match those of the CO(1-0) 
emission in single-dish data. The [NII]/H$\alpha$, [SII]/H$\alpha$ and [OI]/H$\alpha$ ratios vary little with 
position, local H$\alpha$ surface brightness or between clusters. 

We propose that the H$\alpha$ and CO emission arise in molecular clouds heated by a starburst, and that the latter 
has been triggered by interaction with a secondary galaxy. Such CO emission is known to trace massive ($>10^{10}$\Msun) compact ($<20$\kpc) 
reservoirs of cool molecular gas, and it is plausible that an infalling galaxy would disturb this gas, distorting the 
H$\alpha$ morphology and initiating widespread star formation. We also examine the role of cloud-cloud 
collisions in the undisturbed molecular gas reservoir, and suggest that they might be an important source of excitation for the emission line
gas in the cores of lower H$\alpha$ luminosity clusters with less intense star formation.
\end{abstract}


\section{INTRODUCTION}
In recent years our understanding of the X-ray cooling flow phenomenon in galaxy cluster cores
has been revolutionised. Throughout most of the 1980s and 1990s, X-ray 
observations suggested that gas in the central 100\kpc~is cooling out at rates of up to several 
hundred solar masses per year, but the lack of evidence for a reservoir of cooled gas led to heated 
debate (summarised by Fabian~1994) over this interpretation of the X-ray data. Results from 
{\em XMM-Newton} and {\em Chandra} have since led to a sharp downward revision in X-ray cooling 
rates (e.g. Schmidt, Allen \& Fabian 2001) and also reveal a strong deficit of line emission from 
gas cooling below $T_{\rm{virial}}/3$ (Peterson et al.~2003). The implication is that X-ray cooling is quenched, for which 
numerous mechanisms have been proposed, including: rapid mixing of hot and cold phases, 
inhomogeneously distributed metals in the intracluster medium (Fabian et al.~2001,2002); active galactic 
nucleus (AGN) heating by jets (Br\"{u}ggen \& Kaiser 2003) and sound waves (Fabian et al. 2003); thermal 
conduction of heat from the hotter outer parts of the cluster into the cooler core (Voigt et al. 2002); 
a significant relativistic cosmic ray component frozen into the thermal gas (Cen 2005); the release 
of gravitational energy from blobs of gas which detach from the bulk flow and fall directly into the 
core (Fabian 2003).

Concurrently, significant progress has been made in identifying cool gas and dust in cluster cores. 
Edge~(2001) detected CO emission in the centres of 16 cooling flows, consistent with $10^{9}-10^{11.5}$\Msun~of \H2~at 20--40\K~for a standard CO:\H2~conversion (see also Salom\'e \& Combes 2003). These are roughly the 
masses expected, given the revised cooling rates and likely ages. Interferometry shows further that 
the CO emission is localised within the central few arcsec of the cluster (Edge \& Frayer 2003; Salom\'e \& Combes 2004). The frequent occurrence of smaller masses ($\sim 10^{5}-10^{6}$\Msun) of hot \H2~has also been established (e.g. Edge et al.~2002; Jaffe, Bremer \& van der Werf~2001), and excitation analysis suggests that this hot \H2~is a high pressure, transiently-heated component (Wilman et al.~2002). Both CO and \H2~emissions correlate well with the strength of the H$\alpha$ emission from ionized gas at $10^{4}$\K, whose prevalence in these environments, often in the form of spectacular filaments, has long been known (e.g. Hu et al.~1983; Crawford et al.~1999). Despite the clear association between optical line emission and short central X-ray cooling times (Peres et al.~1998; Bauer et al.~2005), their physical relationship is ill-understood. Photoionisation by the observed excess population of hot massive stars can energetically account for the H$\alpha$ luminosities in the most luminous systems (Allen~1995; Crawford et al.~1999). {\em Spitzer} MIPS photometry of 11 CCGs by Egami et al.~(2006) also shows that the most H$\alpha$-luminous in their sample (A1835, A2390 and Zw3146) have prominent far-infrared thermal dust emission plausibly powered by star formation, two of them with $L_{\rm{IR}} > 10^{11}$\Lsun. At lower H$\alpha$ luminosities the picture is less clear: the tapping of energy from the intracluster medium (ICM) through turbulence (Crawford \& Fabian~1992) and heat (Sparks et al. 2004) are just two mechanisms which have been invoked to explain the optical nebulosity in such systems.

In this paper we present integral field spectroscopy of the ionized gas in the cores of four such clusters, A1664, A1835, A2204 and Zw8193. The principal aim is to obtain a full two dimensional view of the distribution and kinematics of the gas through high resolution spectroscopy of the H$\alpha$+[NII] emission line, with additional ionization information being gleaned from the [SII]$\lambda \lambda$6717,6731 and [OI]$\lambda \lambda$6300,6363 lines where possible. These four central cluster galaxies (CCGs) all have H$\alpha$ luminosities exceeding $10^{42}$\ergps, making them 4 of the top 6 most-H$\alpha$ luminous systems in the extensive CCG spectroscopic survey by Crawford et al.~(1999). In this regime of H$\alpha$ luminosity, photoionisation by a young stellar population can account energetically for the luminosity of the H$\alpha$ nebulosity (Allen~1995; Crawford et 
al.~1999). In addition to an analysis of the CCGs, we also present spectroscopy of other sources within the IFU field of view, including other cluster galaxies and (in the case of A2204) a serendipitous gravitationally-lensed background galaxy. We first present results for the clusters individually and then summarise and interpret their generic features. Throughout the paper we assume a cosmology with $H_{\rm{0}}=70$\kmpspMpc, $\Omega_{\rm{M}}=0.3$ and $\Omega_{\rm{\Lambda}}=0.7$ and all physical quantities quoted from other papers have been converted accordingly.

\section{OBSERVATIONS AND DATA REDUCTION}

\subsection{VLT-VIMOS Integral Field Spectroscopy}
The observations of A1664, A1835 and A2204 were taken in service mode on 2003 April 11 with the integral 
field unit (IFU) of the Visible Multiobject Spectrograph (VIMOS) on UT3 of the 8.2m Very Large Telescope 
(VLT) at ESO Paranal in Chile (for further information on VIMOS see LeFevre et al.~2003). They were among the 
first taken with the VIMOS IFU in open time. The IFU was operated in HR-Red mode, offering a 
$27\arcsec \times 27$\arcsec~field of view covered by 1600 optical fibres of 0.67\arcsec~diameter. The 
fibres are coupled to a microlenses to ensure near-continuous sky coverage. The field of view in this IFU mode is split into 4 quadrants 
of 400 fibres, three of which disperse the light with the HR-Red grism over 6250--8700\AA, while the 
fourth quadrant employs the HR-Orange grism spanning 5500--7450\AA. The dispersion and 
spectral resolution are approximately 0.6\AA~per pixel and 1.8\AA~FWHM, respectively. For each of 
the three targets a pair of 1670s exposures was taken, with a pointing dither of $(\Delta \alpha, 
\Delta \delta)= (-6\arcsec, -6\arcsec)$ between them. The seeing was in the range 0.5--1\arcsec~throughout. 
Table 1 provides an observation log.

The data were reduced on site at the National Institute for Astrophysics (INAF) in Milan using the 
VIMOS Interactive Pipeline and Graphical Interface (VIPGI) (Scodeggio et al.~2005) developed by the VIRMOS consortium. 
The raw two-dimensional frames for each quadrant were bias-subtracted, and the spectra for the individual 
fibres were traced, extracted and flat-fielded. The pipeline wavelength calibration was affected by 
instrumental flexure in the time interval between science and 
calibration observations. The wavelength calibration was thus performed manually in IRAF using night-sky emission 
features, rebinning to a common dispersion of 0.6\AA~per pixel. Using the mapping from detector to 
aperture fibre position, the four quadrants were united into a single $40 \times 40$ fibre data 
cube for each pointing, covering the wavelength range 5500--8740\AA~(i.e. large enough to 
accommodate the wavelength range of the HR-red and HR-orange grisms). The raw data format is such that the centres 
of adjacent fibres are separated by 5 pixels on the detector whilst the light from each fibre covers 3 pixels FWHM. 
There is thus a small amount of fibre-to-fibre light leakage implying that the spectra of individual fibres will not be completely 
independent. However, since the seeing is comparable to the fibre size and we are chiefly concerned with diffuse extended emission, 
the effect is not considered important and no corrections for it were made.

Subtraction of the night-sky emission background was performed by fitting gaussian profiles to a 
triplet of sky features at 7316, 7329 and 7341\AA~in each fibre. This intensity information was used 
to correct for fibre-to-fibre transmission variations across the field of view, and then to create a
night-sky emission spectrum from a source-free region which was then subtracted across the whole cube.
Due to the use of different grisms, sky subtraction was performed separately for quadrants 1--3 and 4,
respectively. For A1664 and A2204 the two dithered sky-subtracted datacubes for each target were then 
combined into a single datacube $49\times49$ fibres in size ($33\arcsec  \times 33$\arcsec). For A1835, only the 
second of the two pointings contains significant line emission, since the target was inadvertently placed 
too near the edge of the field of view. Flux calibration of the spectra was not performed, but comparison with published emission 
line slit fluxes (Crawford et al.~1999) suggets that H$\alpha$ emission is detected down to minimum surface brightness levels
in the range 0.5--1.0 $\times 10^{-16}$\emSB. Following the reduction of the raw data, all subsequent cube manipulation and emission line 
analysis were performed within {\sc IDL}.

\subsection{WHT-OASIS Integral Field Spectroscopy}
Observations of the redshifted H$\alpha$ emisson in Zw8193 were made with the
OASIS IFU on the William Herschel Telescope (WHT) on La Palma on 2005 September 2 in 
0.8\arcsec~seeing and photometric conditions. We used the MR\,807 filter which covers a wavelength
range of 7690 - 8460\AA~at a spectral resolution of
$\lambda$/$\Delta\lambda$=2020 (2.2\AA~per pixel).  The 7.4\arcsec$\times$10.3\arcsec~field of
view was sampled contigiously with 0.26\arcsec~fibres. The total integration time was 
7200s, split into 4x30mins and each dithered by one IFU lenslet to account for bad pixels. For further information
about the OASIS IFU at the WHT see Benn et al.~(2003).

The OASIS data were reduced using the {\sc xoasis} data-reduction package
(Rousset 1992) developed at CRAL (Lyon).  The reduction steps include bias
and dark subtraction, extraction of the spectra using an instrument model,
wavelength calibration, cosmic-ray rejection, low frequency flatfielding
and sky-subtraction.  The individual data-cubes were then mosaiced (taking
into account the small spatial offsets between exposures) and resampled
onto a common grid of 0.26\arcsec~$\times$0.26\arcsec~to create the final datacube. 
The limiting emission line surface brightness is comparable to that reached in the VIMOS observations.
Line fitting to the OASIS data was performed by binning up $3 \times 3$ pixels in order to better match the 
seeing.

\begin{table}
\caption{Summary of the observations}
\begin{tabular}{|llllllll|} \hline
Target  & z &  IFU & Exposure & Seeing  & Scale \\
        &   &  & time (s) & (arcsec)& kpc/arcsec \\ \hline
A1664	& 0.1276 & VIMOS & 3340 & 0.5--1.0 & 2.3 \\ 
A1835	& 0.2523   & VIMOS  & 3340 &     0.5--1.0 & 3.9  \\
A2204   & 0.1514   &  VIMOS & 3340 &     0.5--1.0 & 2.6 \\ 
Zw8193  & 0.1825 & OASIS & 7200 & 0.8 & 3.1 \\ \hline
\end{tabular}
\end{table}

\section{A1664}

\subsection{H$\alpha$ morphology and kinematics}
With a extinction-corrected H$\alpha$ slit luminosity of $1.6 \times 10^{42}$\ergps, the central cluster galaxy of A1664 is 
among the brightest H$\alpha$ emitters in the optical spectroscopic survey of the ROSAT Brightest Cluster 
Sample (BCS; Crawford et al.~1999); fits to the excess blue continuum light imply a visible star formation rate
of 23\Msunpyr~(see also Allen~1995). 

The complexity of the line emission in A1664 is evident in the reduced two-dimensional frame of one 
of the quadrants, shown in Figure 1. The emission spans a large range in velocity and at many spatial locations two velocity components are required to satisfactorily model the H$\alpha$+[NII]$\lambda\lambda$6548,6584 profiles. Within each spatial pixel, the lines were fitted with zero, one or two H$\alpha$+[NII]$\lambda\lambda$6548,6584 gaussian velocity components atop a flat continuum level; within each velocity component the H$\alpha$/[NII] was allowed to vary but the [NII]$\lambda\lambda$6548,6584 double ratio was fixed at the theoretical 1:3 ratio. A simple chi-squared fitting criterion was used to decide whether one, two or zero velocity components were fitted at each position. The complex was fitted within the wavelength range 7330--7450\AA~(i.e. 200 dispersion bins) and the noise-level was estimated from the spectrum in the wavelength range 6490--7500\AA. Chi-squared values were evaluated for fits with 0, 1 or 2 velocity components, $\chi^{2}_{\rm{0}}$, $\chi^{2}_{\rm{1}}$ and $\chi^{2}_{\rm{2}}$, respectively. For $\chi^{2}_{\rm{0}} - \chi^{2}_{\rm{1}} > 150$, 1 velocity component was fitted;  for $\chi^{2}_{\rm{1}} - \chi^{2}_{\rm{2}} > 10$, two components were deemed necessary. All fits were examined by eye to confirm whether they were reasonable and to remove spurious fits to obvious noise spikes.

Fig.~\ref{fig:A1664FLUX} shows the H$\alpha$ and continuum maps derived from these emission line fits. For clarity, three separate H$\alpha$ maps are shown: the total flux (i.e. all the H$\alpha$ emission), and the separate red and blue component fluxes for those pixels fitted with two H$\alpha$ velocity components. The velocity and FWHM fields derived from these fits are shown in Fig.~\ref{fig:A1664VEL}, and for clarity are again shown separately for the red and blue components where two components are required, and for those regions where only a single velocity component is needed. Line profile fits are shown in Fig.~\ref{fig:A1664profs} for selected regions of H$\alpha$ emission in Fig.~\ref{fig:A1664FLUX}. From Fig.~\ref{fig:A1664FLUX}, we see that the H$\alpha$ emission has a linear filamentary structure spanning approximately 13 arcsec (31 kpc) and passing through the nucleus of the central cluster galaxy along a north-east -- south-west axis. Two velocity components are detected along its whole length: the red component peaks on the nucleus of the CCG, but the blue component exhibits two maxima (labelled B1 and B2) located approximately 1.5 arcsec from the nucleus, along an axis perpendicular to the general direction of the filament. This filament and nuclear bipolar structure are discussed further below.

\subsection{The 30-kpc H$\alpha$ filament and nuclear bipolar structure} 
The presence of two velocity components along the length of the filament (most clearly at its NE end where complications from the nuclear emission of the CCG are minimal) suggests that it might be due to a disturbed secondary emission component superimposed on a quiescent background, or a stream of gas wrapped around the CCG, or possibly related to some form of outflow. It is plausibly related to the galaxy (obj \#1) situated just beyond the SW extremity of the filament in Fig.~\ref{fig:A1664FLUX}, which is confirmed to be at the cluster redshift (see Fig.~\ref{fig:A1664VEL}) with a line of sight velocity of $-560$\kmps~in the velocity system adopted in Fig.~\ref{fig:A1664VEL}. This gas may have been stripped from or disturbed by obj \#1. Indeed, the velocity of the blue emission component at position B2 ($-450$\kmps) is similar to that of obj \#1 ($-560$\kmps). 

The bipolar structure may instead be an outflow from the CCG driven by starburst activity. The line of sight velocity difference between the blue emission components at B1 and B2 is $\simeq 350$\kmps. This suggests a minimum deprojected outflow velocity of 175\kmps, but the symmetrical location of B1 and B2 about the nucleus suggests that the outflow may be principally directed into the plane of the sky with a much higher velocity. In the latter case, the observed radial extent of the outflow (1.5\arcsec=3.5\kpc) can be used to constrain its age. Assuming a deprojected outflow speed of 600\kmps~as for the H$\alpha$~filaments in M82 (Shopbell \& Bland-Hawthorn~1998), the implied age is $\sim 6$~Myr. By contrast, the orbital timescale of galaxy \#1 responsible for the creation of the filament is $\sim 100$~Myr~(assuming a circular orbit of radius 15\kpc~traversed at 500\kmps). Considering the mis-match in these timescales, it is too coincidental that an outflow from a nuclear starburst triggered by the passage of galaxy \#1~should be observed to have a similar extent to the overall stream of gas. We thus favour an interpretation in which the complex H$\alpha$ structure is due to a disturbance induced by, or the stripping of gas from, galaxy \#1. 

\subsection{Archival VLT-FORS1 V, R and I imaging}
To shed further light on the origin of the complex H$\alpha$ distribution, we retrieved archival VLT-FORS1 imaging of A1664. The dataset 
comprises images three 330 second exposures taken through the V, R, and I filters (Bessel photometric system) on 2001 May 29 in photometric conditions and 0.8\arcsec~seeing. Following bias-subtraction and flat-fielding, we merged them into a true-colour image which is displayed in 
Fig.~\ref{fig:A1664FORS1}. H$\alpha$ contours from the VIMOS IFU observations were aligned and overlaid on this image.

The main features apparent from Fig.~\ref{fig:A1664FORS1} are that: (i) the north-eastern quadrant of the CCG is redder than other parts of the CCG, suggesting dust obscuration; (ii) the south-western quadrant of the CCG is bluer than other cluster galaxies, suggesting more recent star formation; (iii) the red H$\alpha$ velocity component peaks on a nuclear dust lane which bisects the two central peaks of the blue H$\alpha$ velocity component; (iv) the southern extension of the H$\alpha$ emission (labelled B4 in Fig.~\ref{fig:A1664FLUX}) coincides with a blue knot of continuum emission. 

V--R and R--I colours (expressed relative to the colours of the nearest A1664 cluster elliptical) were extracted for regions B1,2,4 and object \#1 in Fig.~\ref{fig:A1664FLUX}. These show that (i) galaxy \#1 has the same colour as other cluster ellipticals at this redshift; (ii) the CCG is $\simeq 0.1$~mag bluer in $V-R$ and $\simeq 0.1$~mag redder in $R-I$ and that the intrinsic dust extinction increases across it from south-west to north-east. From long-slit spectra, the intrinsic extinction inferred from the Balmer decrement is in the range 0.46--0.63 mag (Crawford et al.~1999; Allen~1995). Corrected for this extinction, the same authors show that approximately 90 per cent of the (rest-frame) 4500\AA~continuum is due to O5 and B5 stars giving rise to the 23\Msunpyr~starburst. The gradient of increasing dust extinction from sout-west to north-east could be due to the gradient in time elapsed since the passage of galaxy \#1, which may have triggered star formation and dust production. U or B-band imaging would more clearly show the distribution of recent star formation.


\subsection{Ionization properties}
Although the VIMOS data contain the emission lines of [SII]$\lambda \lambda$6717,6731 and [OI]$\lambda\lambda$6300,6363 in addition to [NII]+H$\alpha$, we do not present an exhaustive study of the ionization properties of the gas. Instead, we impose the kinematics found for [NII]+H$\alpha$ and accordingly fit the [SII] and [OI] doublets with zero, one or two velocity components with only the line intensities allowed to vary, i.e. we do not search for emission in these lines where there is no [NII]+H$\alpha$ detection. In so doing, we discover that the ionisation properties of the H$\alpha$ gas are quite uniform, with no notable spatial gradients. In Fig.~\ref{fig:A1664ions} we show scatter plots of line ratios versus H$\alpha$ flux for the combined sample of lenslets. There is essentially no variation in [NII]$\lambda 6584$/H$\alpha$ and 
[OI]$\lambda 6300$/H$\alpha$ over more than one decade in H$\alpha$ surface brightness.  These line ratios are consistent with the integrated slit values found by Crawford et al.~(1999). The uniformity of the ionization state across the source supports our interpretation that all the gas (the filament and the B1/B2 nuclear structure) has the same origin. For comparison, the emission line filaments in the kiloparsec-scale superwind outflow in M82 have [NII]$\lambda 6584$/H$\alpha=0.3-0.6$ 
(Shopbell \& Bland-Hawthorn~1998).

\subsection{Comparison with CO and H$_{\rm{2}}$ observations}
In the CO line survey of cooling flow CCGs by Edge~(2001), A1664 exhibits unusually broad CO(1-0) emission spanning 620\kmps~full-width at zero intensity, with a total inferred mass of cool molecular hydrogen is $4.4 \pm 0.7 \times 10^{10}$~M\sun. Considering the 23.5~arcsec beam size for these IRAM 30-m observations, its kinematics are a close match to those of H$\alpha$; in particular, the skewness of the CO line towards negative velocities suggests that the blue H$\alpha$ velocity components may have associated CO. A direct comparison is shown in Fig.~\ref{fig:A1664COHa}: the CO(1-0) profile is taken from Edge~(2001), and the H$\alpha$ profile is formed by integrating over the whole source and adding synthetic gaussians for the H$\alpha$ emission only, using the parameters derived from the multi-component fits to H$\alpha$+[NII]. This procedure is used in order to compare the kinematics without complications from the blended [NII] lines. Fig.~\ref{fig:A1664COHa} demonstrates that the widths and centroids of the H$\alpha$ and CO(1-0) emission are indeed comparable, and that much of the breadth of these lines can be attributed to the regions of kinematically disturbed H$\alpha$ emission with two velocity components.

For comparison, in the K-band UKIRT CGS4 spectrum of Edge et al.~(2002), the width of the Pa$\alpha$ line is $\sim 1100$\kmps~FWHM, for a N-S slit of $5.5 \times 1.2$~arcsec. This is consistent with the observed H$\alpha$ kinematics when the VIMOS data are extracted over the same area and fitted with a single velocity component. In contrast, the H$_{\rm{2}}$~v=1-0~S(3) line is not resolved, suggesting an intrinsic width much below the instrumental broadening of 570\kmps~FWHM. This suggests that the hot H$_{\rm{2}}$ is distinct from the CO and H$\alpha$-emitting gas in the case of A1664.

\subsection{Cluster X-ray emission}
A1664 has been observed with the ACIS-S instrument on the {\em Chandra} X-ray observatory for 10\ks. 
In Fig.~\ref{fig:A1664chandra} we show the smoothed 0.5-5\keV~image (courtesy of A.C.~Fabian).
 This shows that the X-ray emission is also elongated along a NE-SW axis with a possible filamentary structure, 
resembling the H$\alpha$ emission but on a much larger spatial scale. A full analysis of the X-ray dataset will
be presented elsewhere. The observed similarity in H$\alpha$ and X-ray morphology is analogous to that seen in 
the core of A1795 (Crawford, Fabian \& Sanders~2005), although in that cluster the emission line gas in the filament is very quiescent, in stark contrast to the case of A1664. As Fig.~\ref{fig:A1664chandra} demonstrates, the CCG's small neighbour galaxy (labelled object \#1 in Fig.~\ref{fig:A1664FLUX}) appears to coincide with a local 
minimum in the X-ray emission.

\begin{figure*}
\includegraphics[width=0.88\textwidth,angle=0]{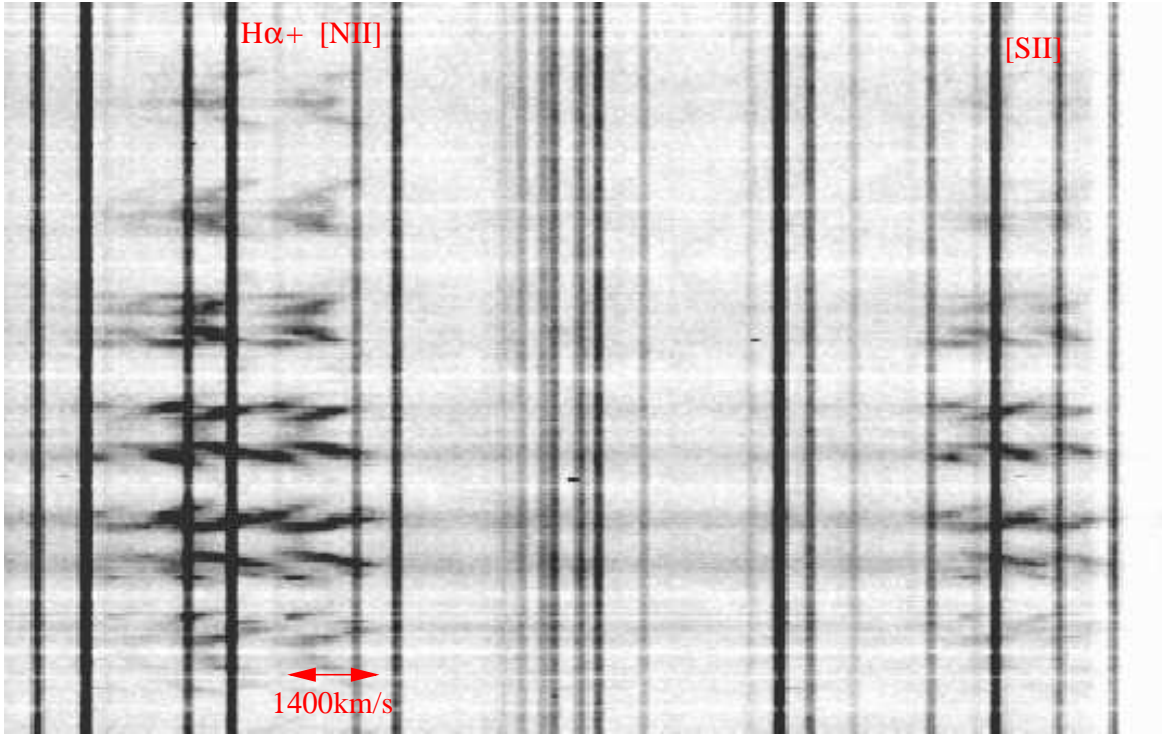}
\caption{\normalsize Part of the reduced 2-D frame for a single quadrant of the A1664 observation. 
The complexity and large velocity range of the emission in H$\alpha$+[NII]$\lambda\lambda$6548,6584 and [SII]$\lambda\lambda$6717,6731 are clearly evident. Note that neighbouring fibres on the sky do not necessarily lie in adjacent rows of this frame.}
\label{fig:A1664_2d}
\end{figure*}

\begin{figure*}
\begin{centering}
\includegraphics[width=14.5cm,angle=0]{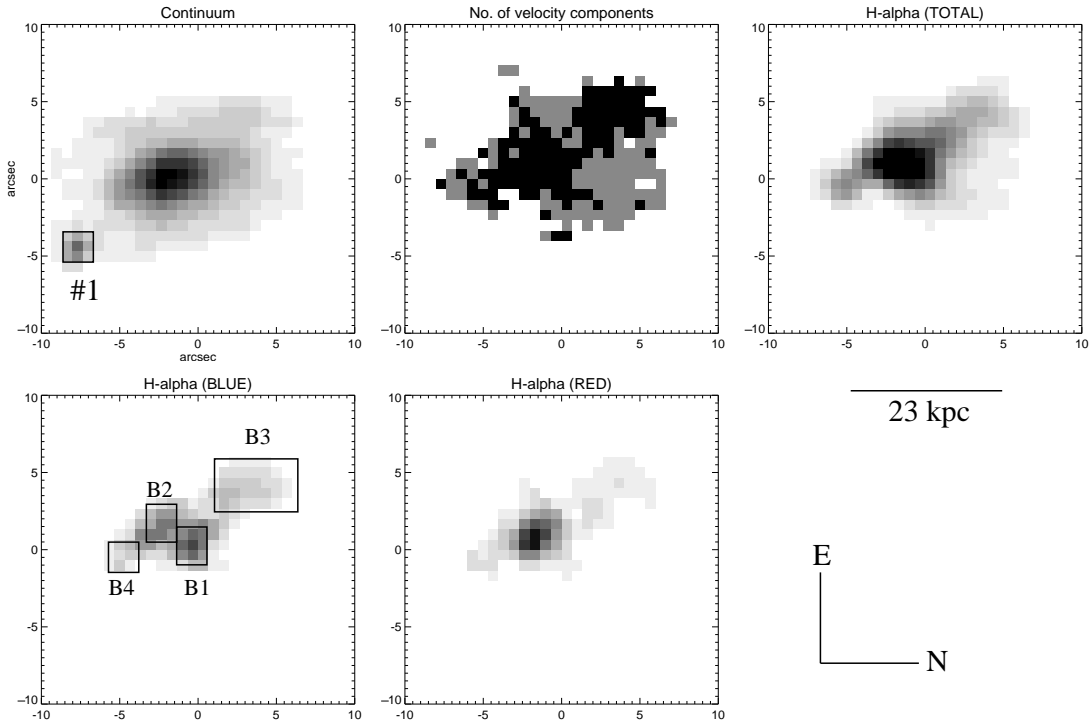}
\caption{H$\alpha$ and continuum (6700--7000\AA~observed frame) intensity plots from multi-component fits to the H$\alpha$+[NII] complex in A1664. These images have been smoothed with a gaussian of FWHM=1.5 pixels (=1 arcsec). The emission line complexes were fitted with 1 or 2 velocity components, shown as grey and black, respectively. The H$\alpha$ (BLUE/RED) panels show the intensities of the two components where fitted. H$\alpha$ (TOTAL) shows the total H$\alpha$ intensity and includes those pixels in which only one velocity component was fitted. The object labelled \#1 in the continuum image is a galaxy at redshift $z=0.1255$ (i.e. in the cluster core), as shown by its spectrum in Fig.~\ref{fig:A1664obj1}. Line profile fits to regions B1--B4 are shown in Fig.~\ref{fig:A1664profs}}
\label{fig:A1664FLUX}
\end{centering}
\end{figure*}

\begin{figure*}
\begin{centering}
\includegraphics[width=4.5cm,angle=0]{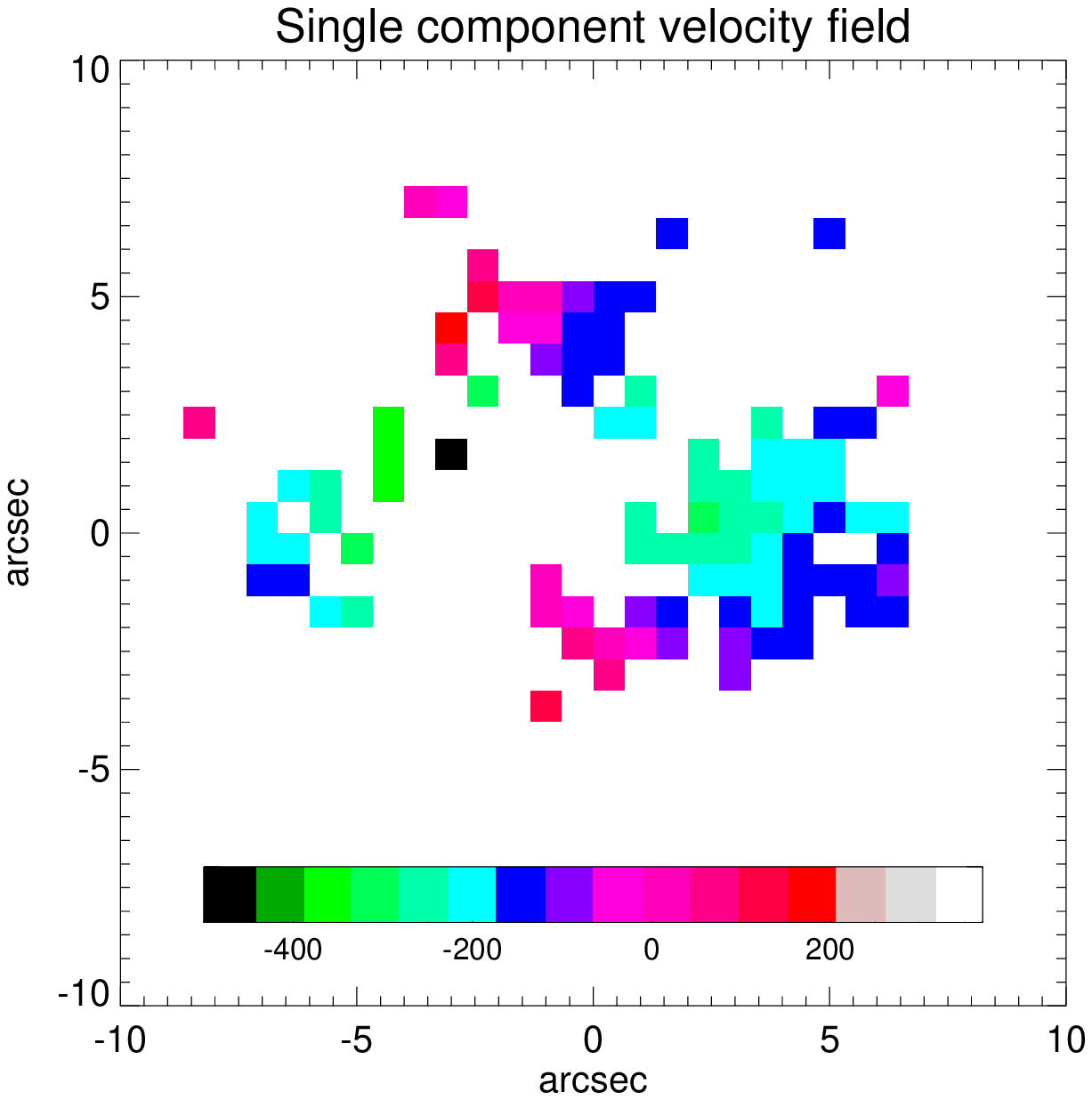}
\includegraphics[width=4.5cm,angle=0]{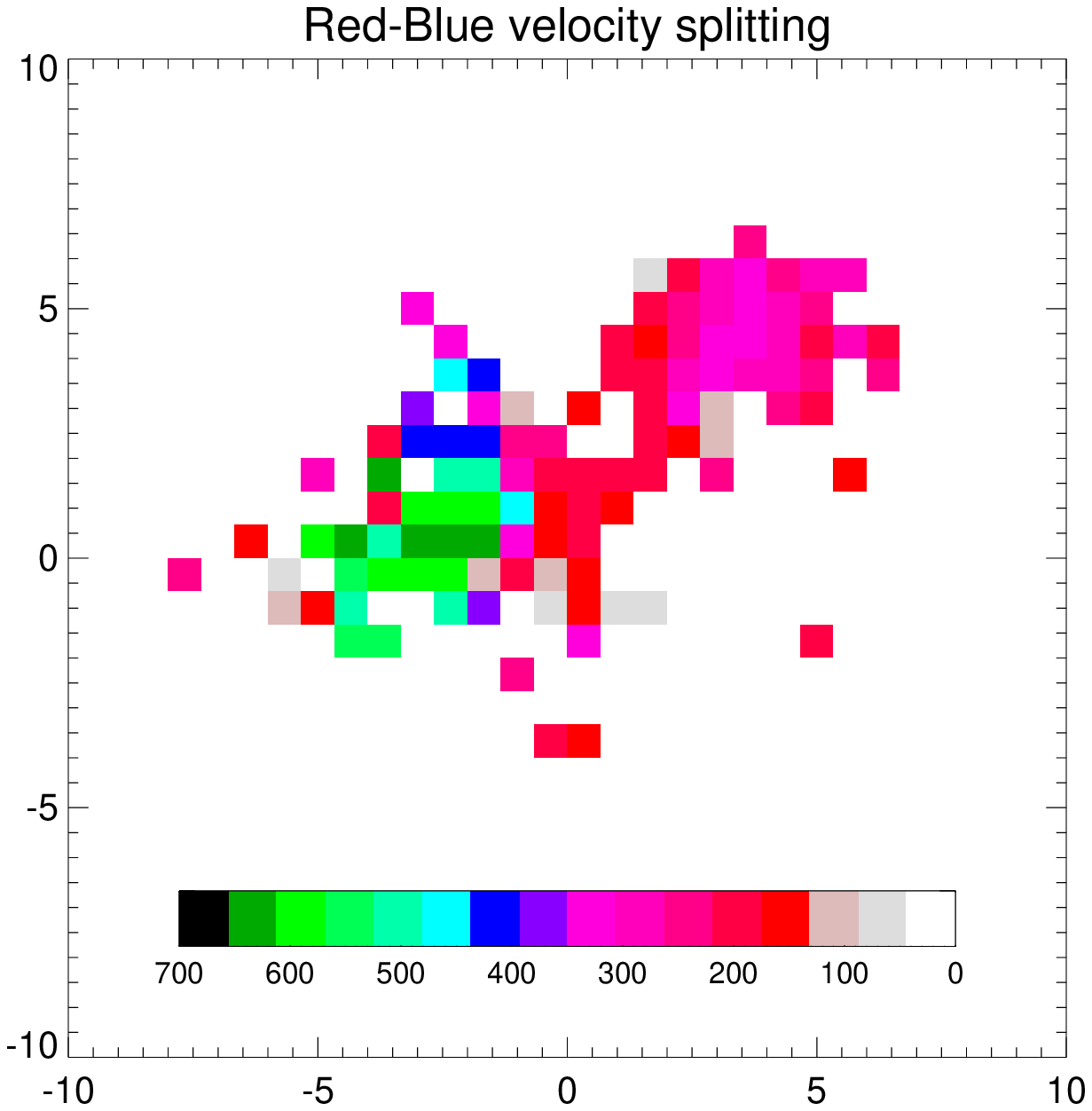}
\includegraphics[width=4.5cm,angle=0]{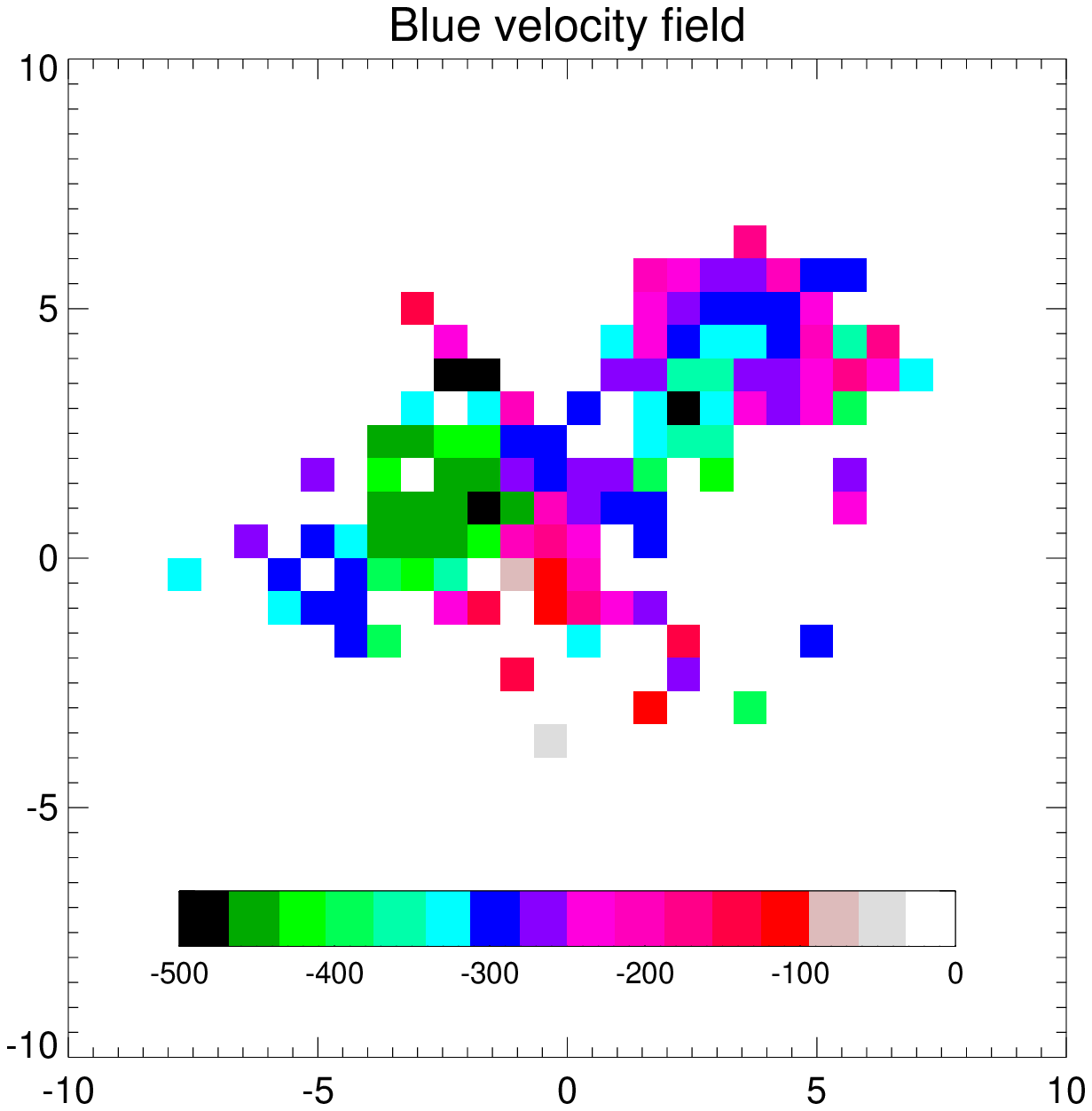}
\includegraphics[width=4.5cm,angle=0]{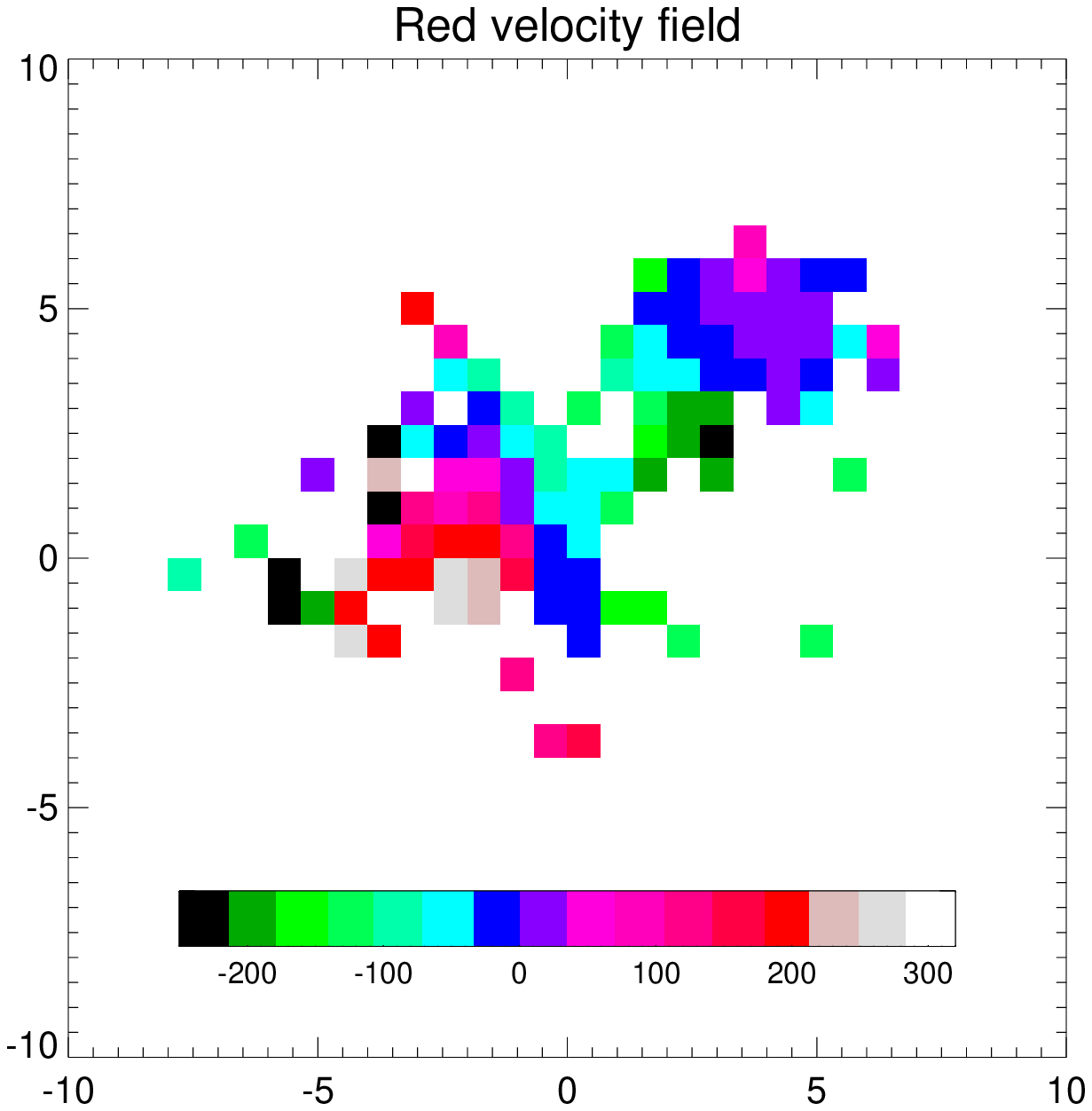}
\includegraphics[width=4.5cm,angle=0]{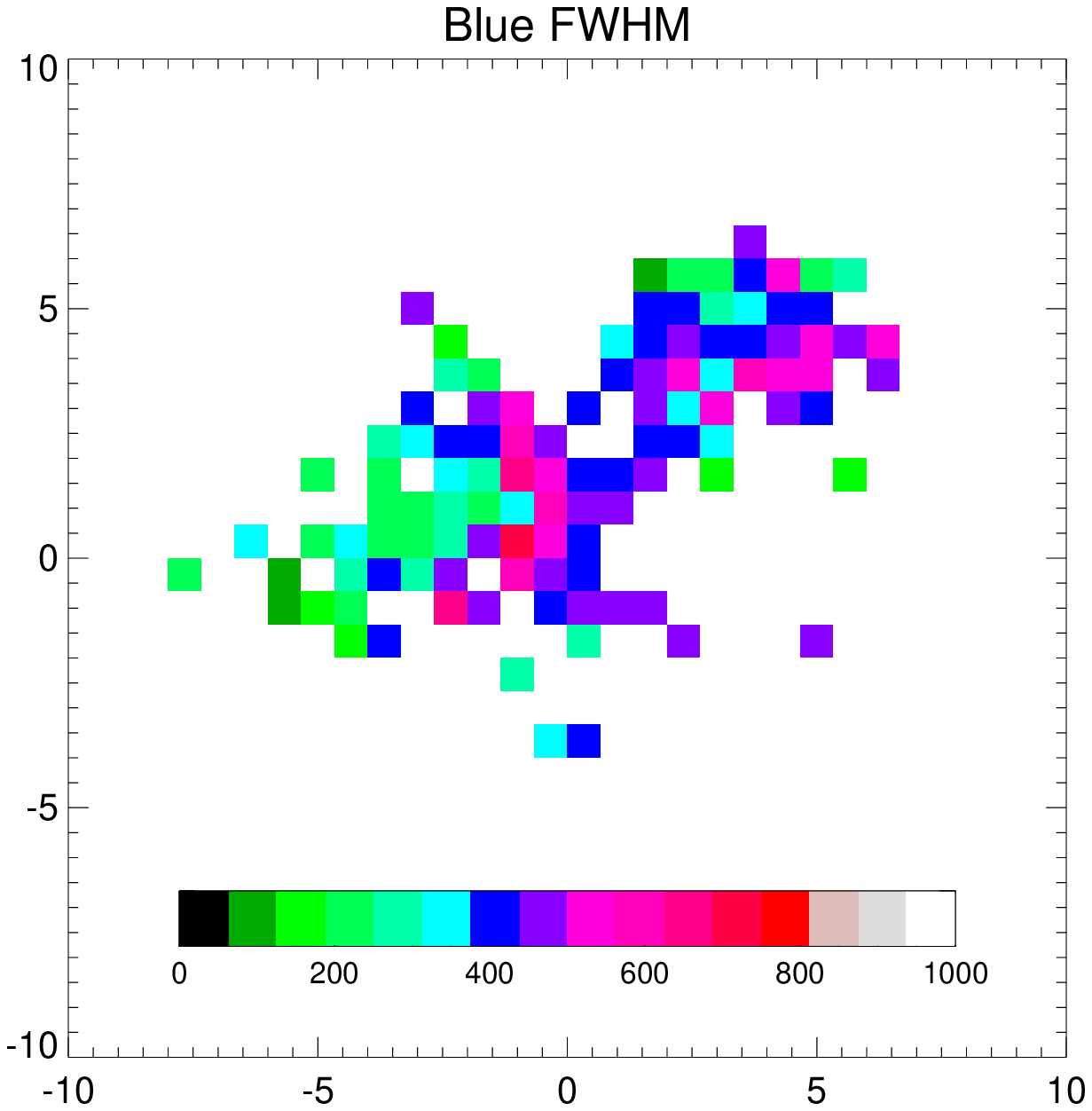}
\includegraphics[width=4.5cm,angle=0]{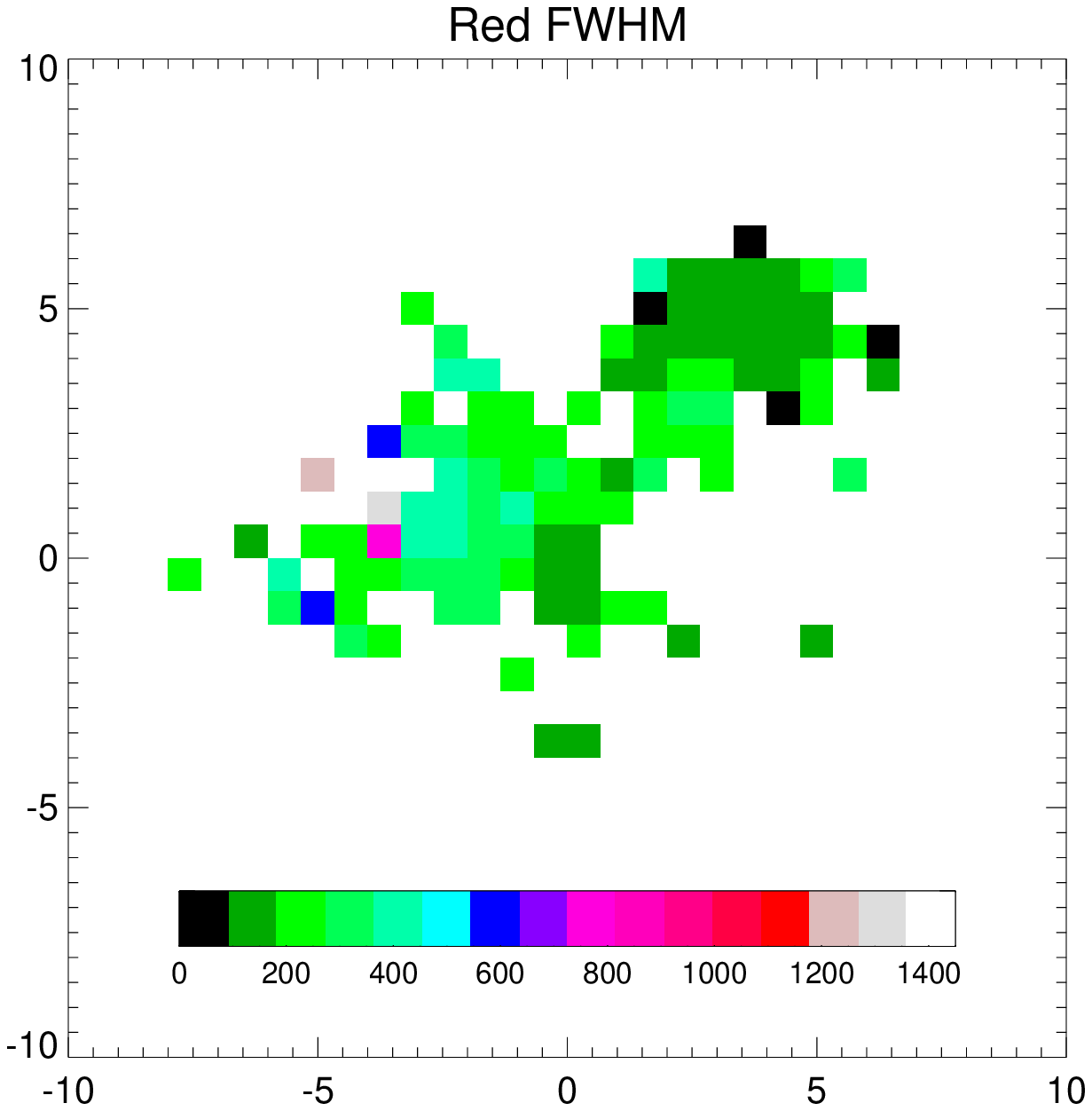}
\includegraphics[width=4.5cm,angle=0]{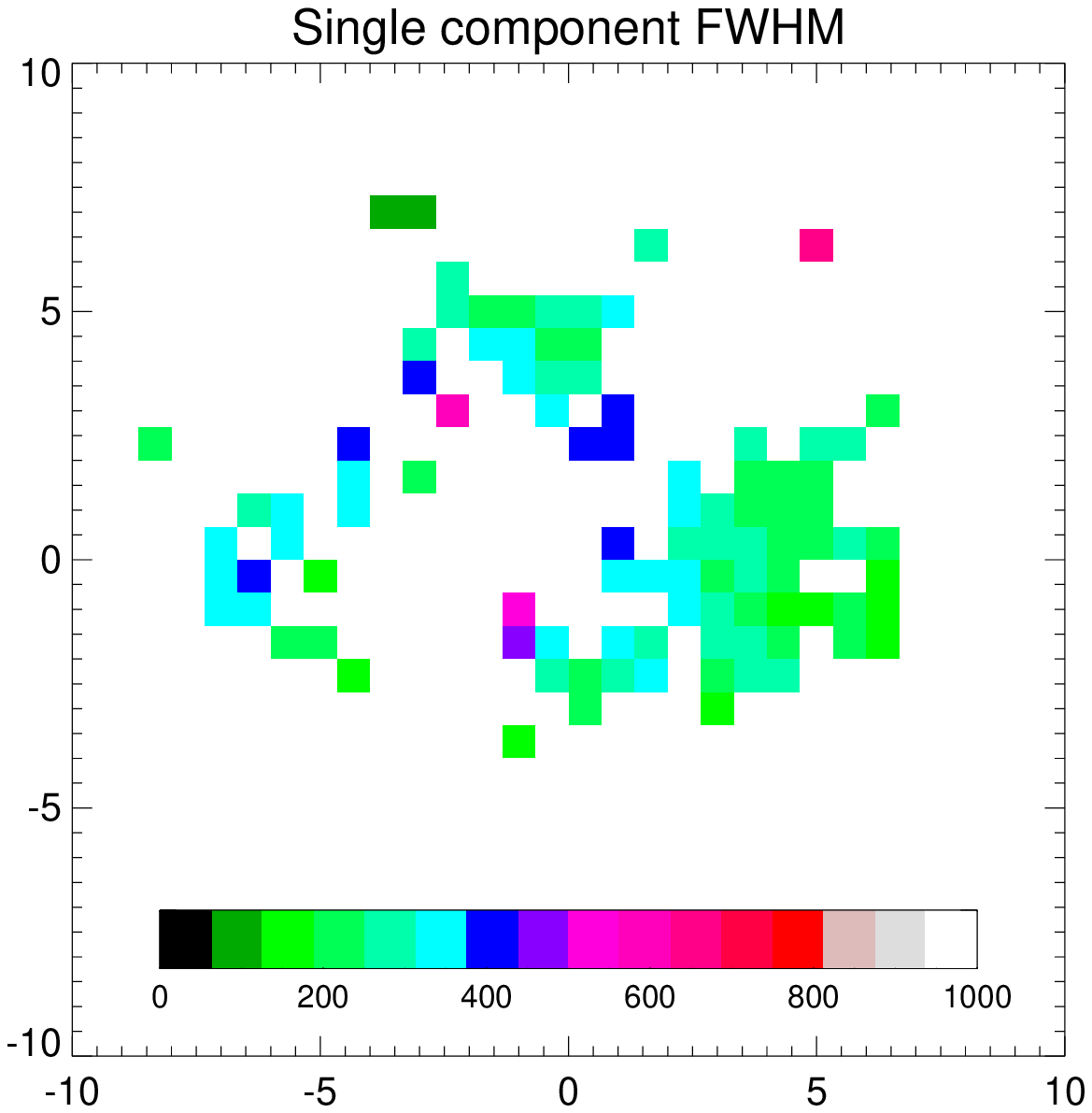}
\includegraphics[width=4.5cm,angle=0]{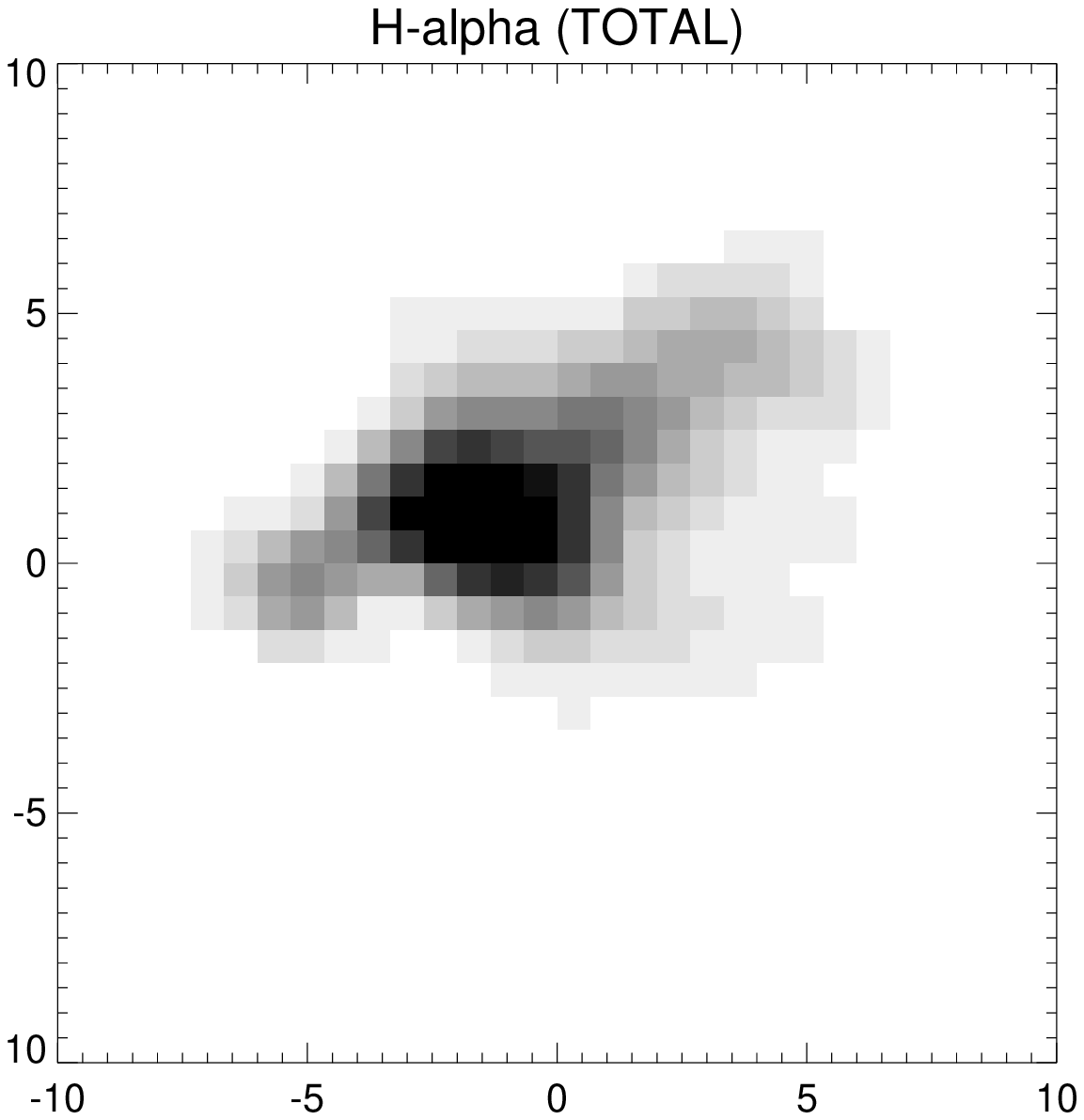}

\caption{H$\alpha$ kinematics as derived from multi-component fits to the H$\alpha$+[NII] complex in A1664. All units are \kmps~and the velocity fields are with respect to a common zeropoint of the nominal cluster redshift ($z=0.1276$). The velocity range spanned by the colour bar varies from panel to panel in order to clearly show the kinematic features of interest. For clarity, the velocity and FWHM fields are shown separately for the red and blue velocity components where two kinematic components were required to fit the line, and for those fibres where just a single velocity component was required to fit it. For comparison, the velocity of continuum object \#1~in Fig.~\ref{fig:A1664FLUX} in this reference system is -560\kmps~(determined from spectrum in  Fig.~\ref{fig:A1664obj1}). For ease of comparison, the total H$\alpha$ intensity map of Fig.~\ref{fig:A1664FLUX} is also shown.}
\label{fig:A1664VEL}
 \end{centering}
\end{figure*}

\begin{figure}
\includegraphics[width=7.5cm,angle=0]{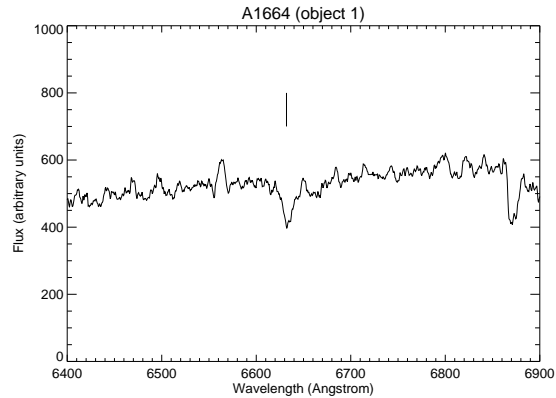}
\caption{The spectrum of continuum object \# 1 in Fig.~\ref{fig:A1664FLUX}. The absorption feature due to the Na D $\lambda 5892.5$ blend is labelled, from which a redshift of $z=0.1255$ is determined. The feature at 6860--6880\AA~is telluric absorption.}
\label{fig:A1664obj1}
\end{figure}

\begin{figure}
\includegraphics[width=9.5cm,angle=0]{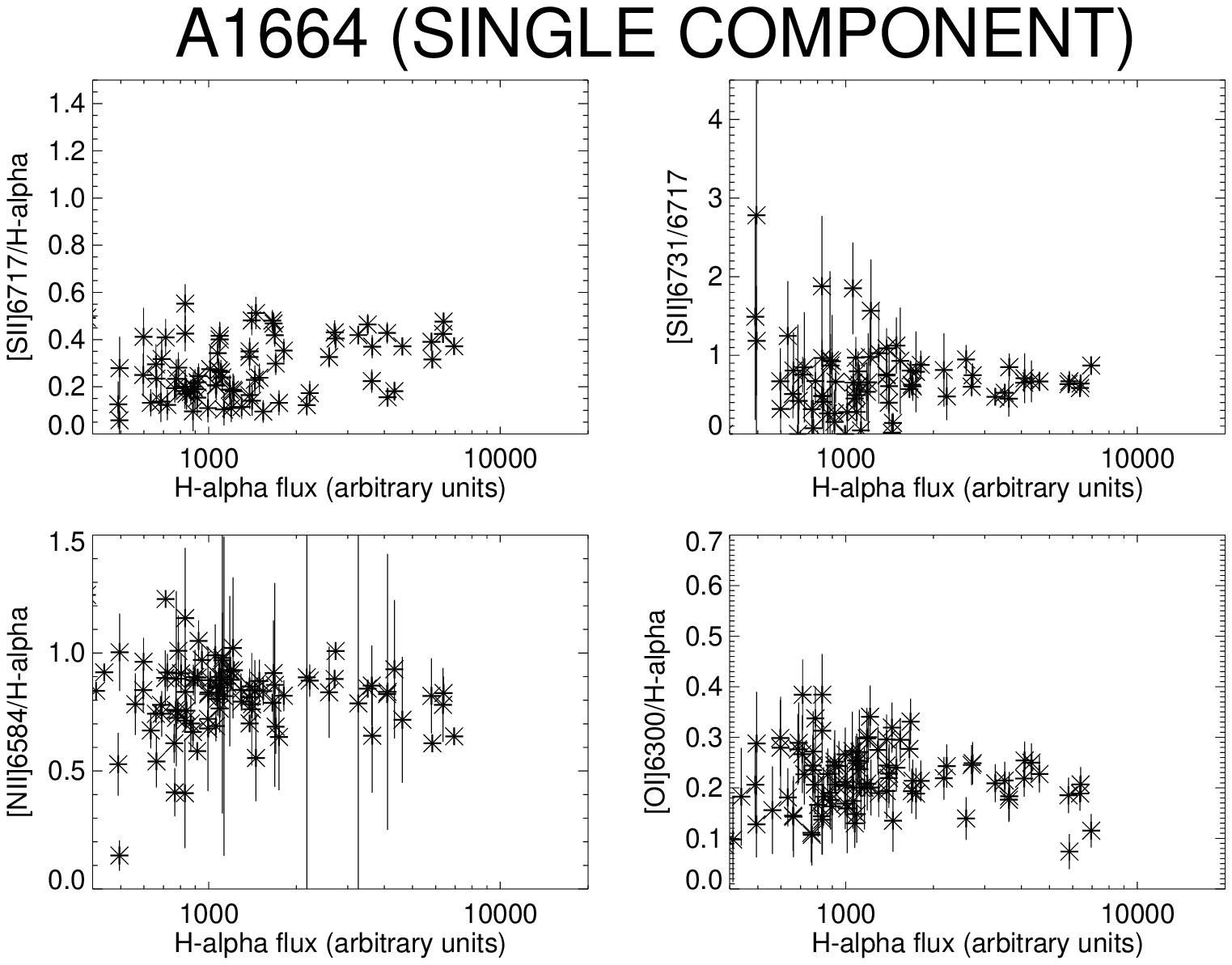}
\includegraphics[width=9.5cm,angle=0]{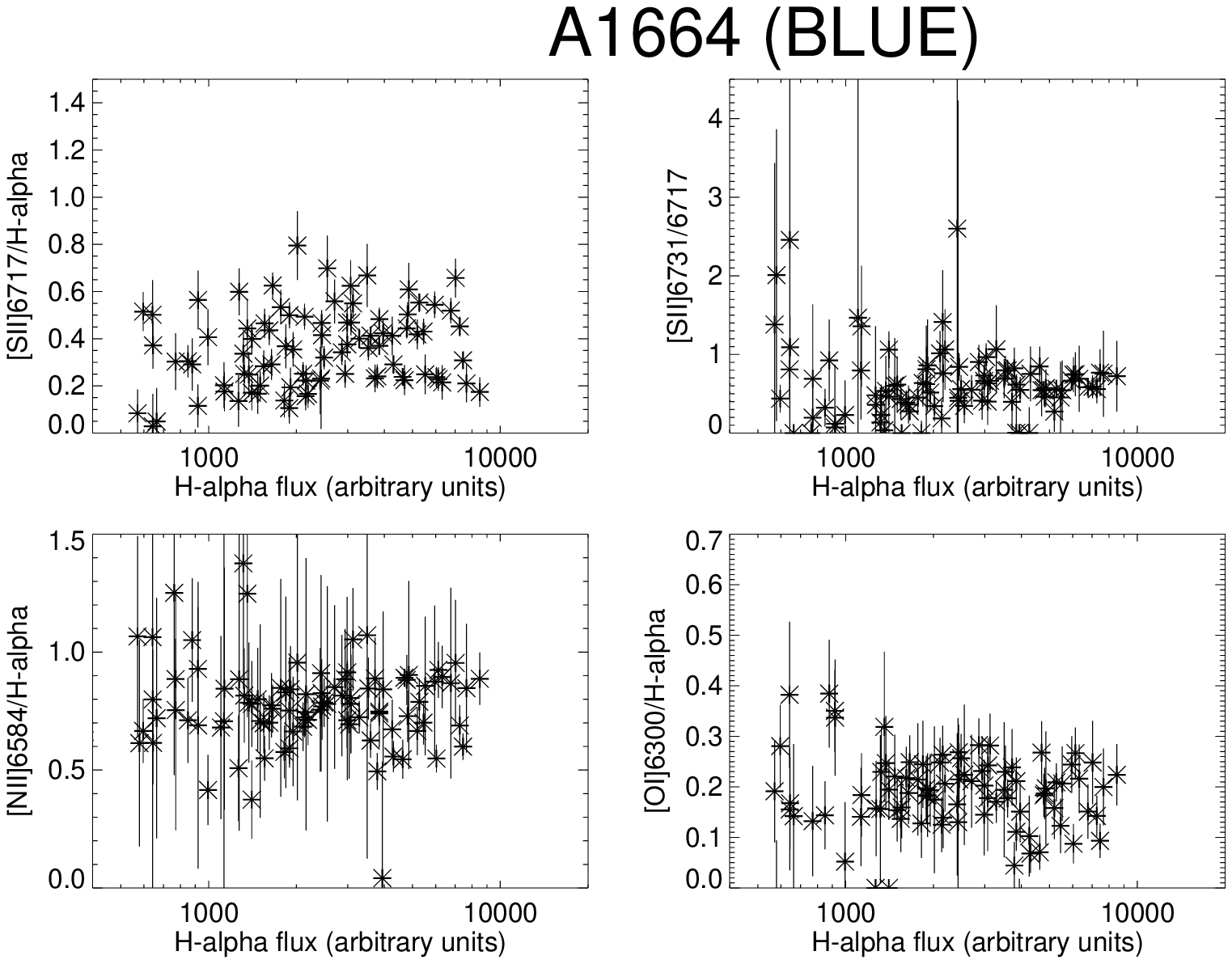}
\includegraphics[width=9.5cm,angle=0]{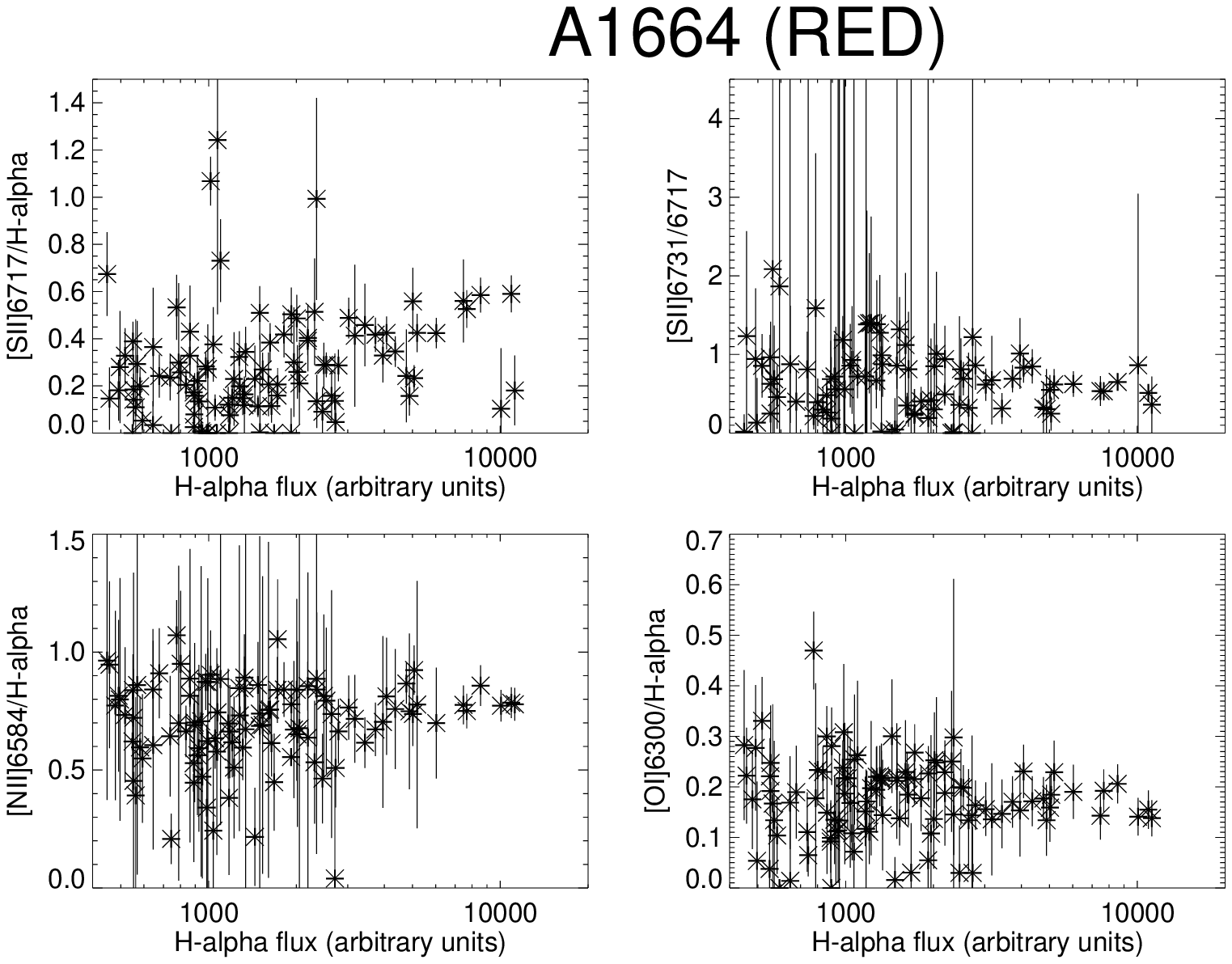}
\caption{Ionisation diagnostic line ratios plotted as a function of H$\alpha$ flux for individual fibres in A1664. Results are 
shown separately for the red and blue velocity components (where fitted), and also for fibres where a single velocity component
was fitted.}
\label{fig:A1664ions}
\end{figure}

\begin{figure}
\includegraphics[width=8.5cm,angle=0]{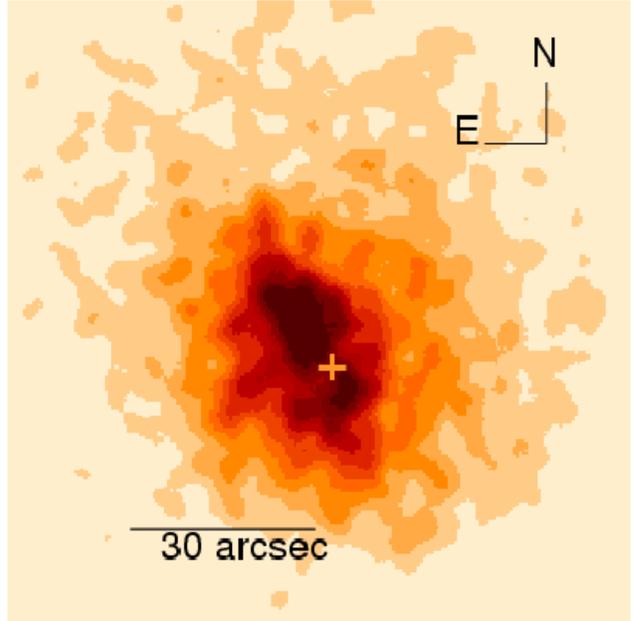}
\caption{The 10\ks~{\em Chandra} ACIS-S image of A1664 in the 0.5--5\keV~energy band with 1.5\arcsec~gaussian smoothing (courtesy A.C. Fabian). The orange cross-hair denotes the position of the galaxy \#1 (Fig.~\ref{fig:A1664FLUX}) close to the CCG.}
\label{fig:A1664chandra}
\end{figure}

\begin{figure*}
\begin{centering}
\includegraphics[width=7.cm,angle=0]{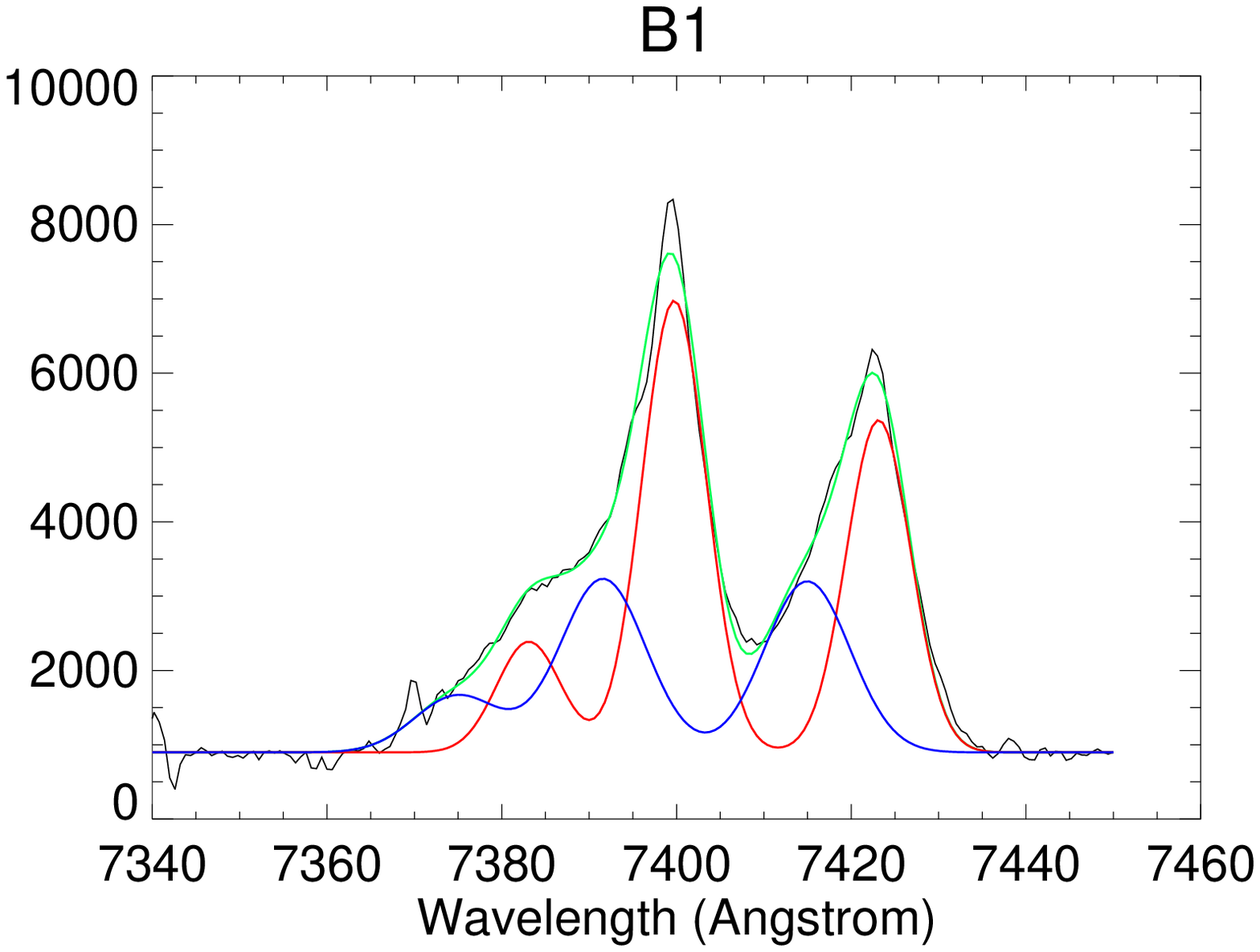}
\includegraphics[width=7.cm,angle=0]{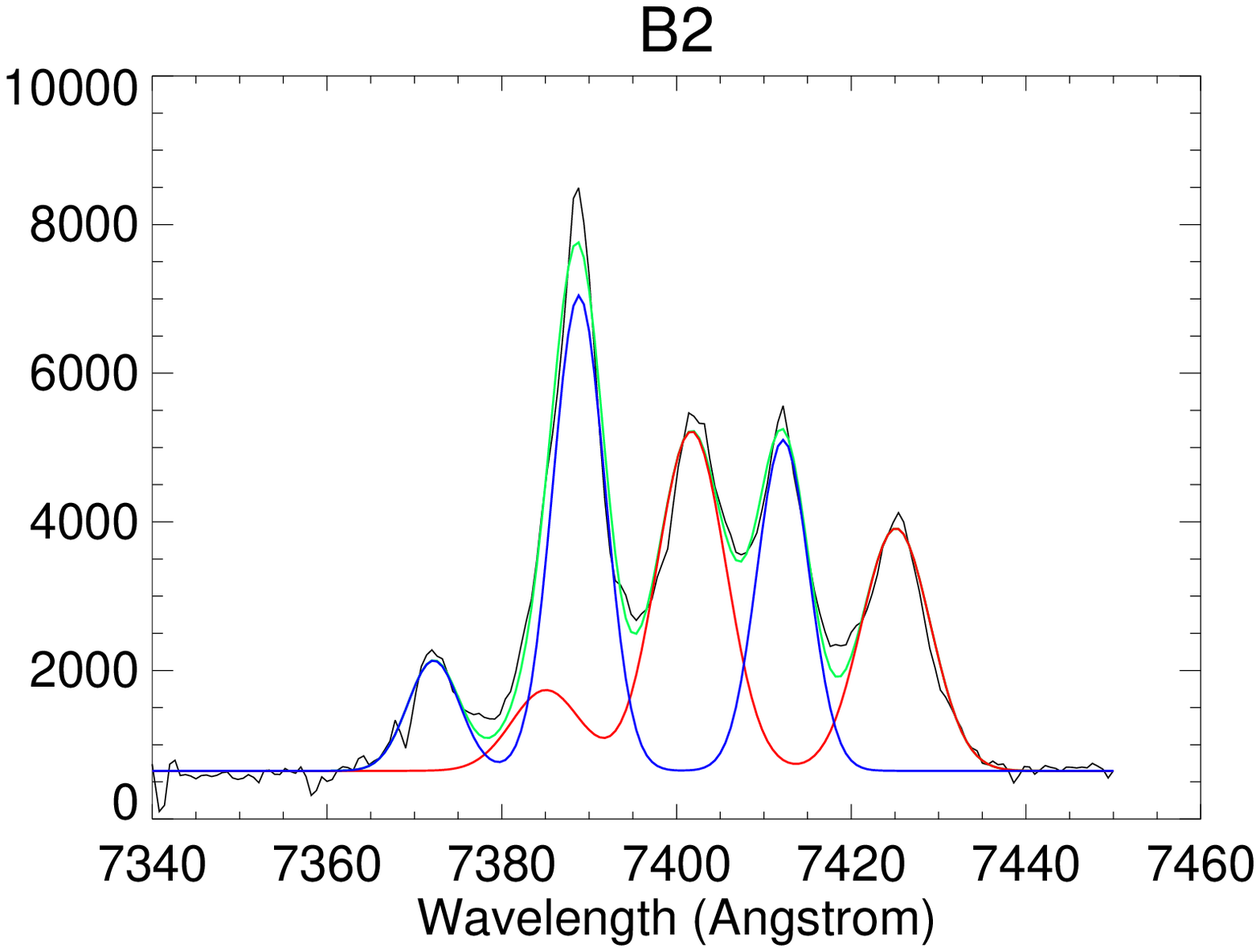}
\includegraphics[width=7.cm,angle=0]{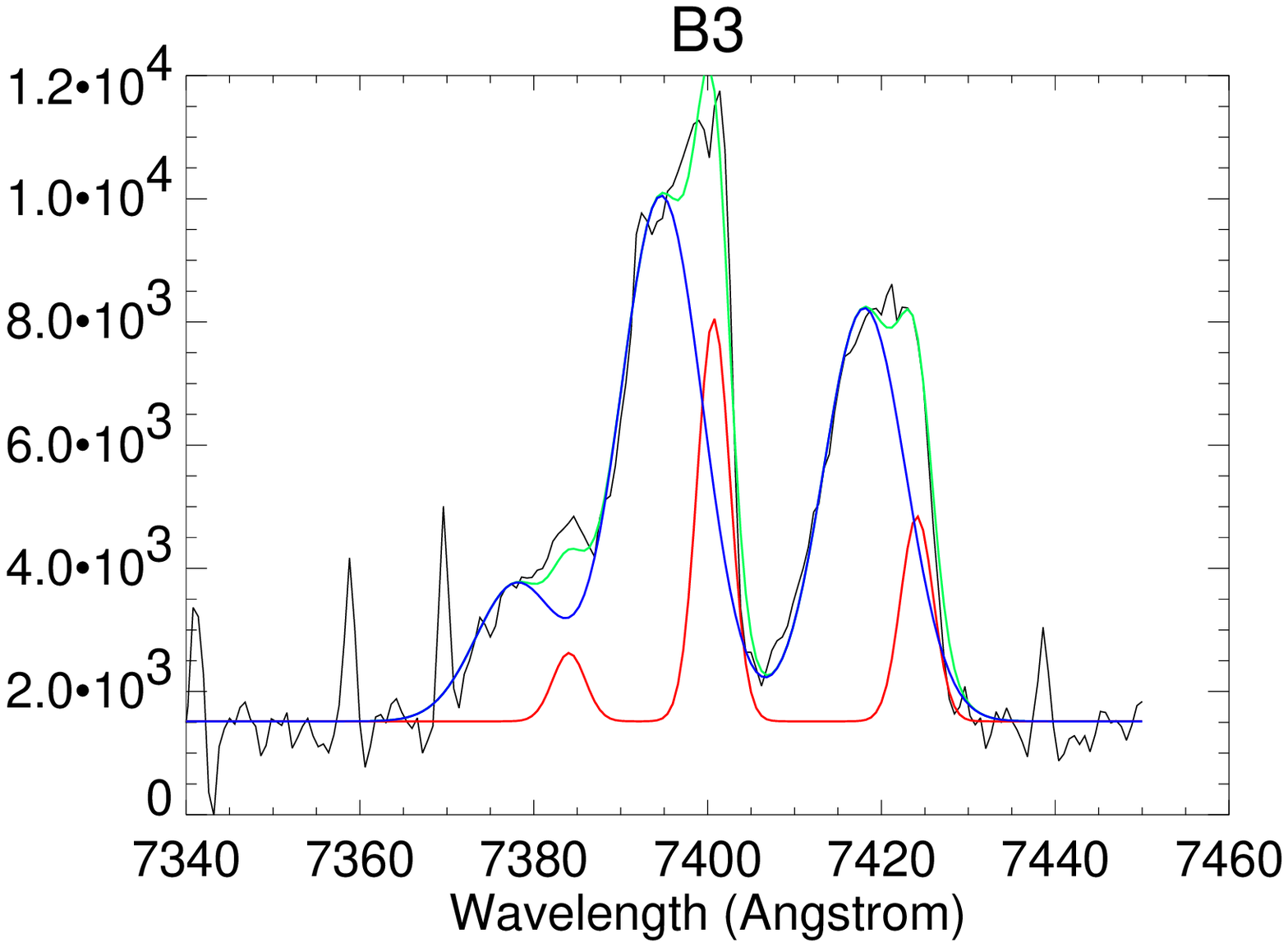}
\includegraphics[width=7.cm,angle=0]{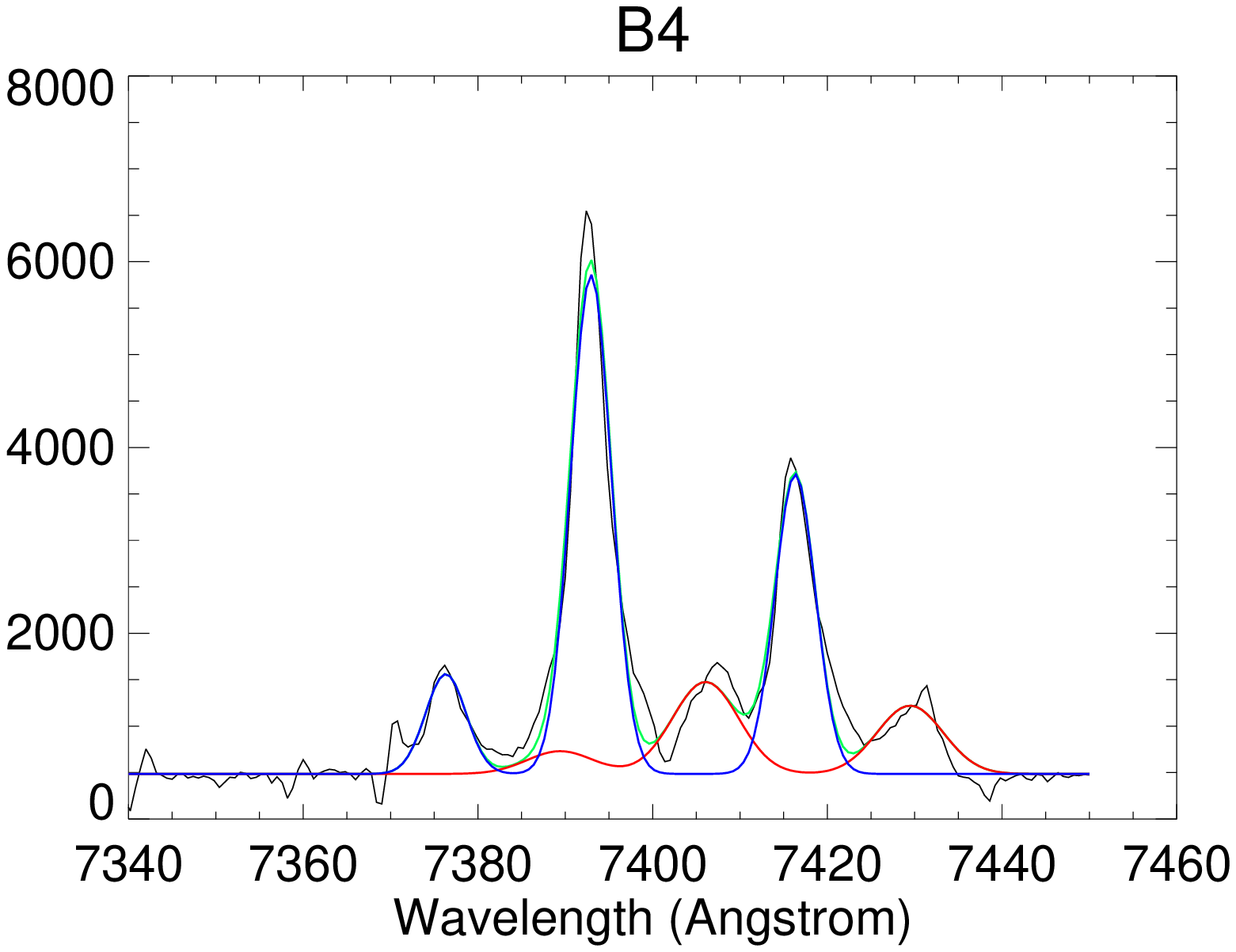}
\caption{Fits to the H$\alpha$+[NII] complex in A1664 for the regions labelled in Fig.~\ref{fig:A1664FLUX}. The two velocity components are shown in red and blue and the green line denotes their sum. Spikes adjacent to the B3 profile are sky emission feature residuals.}
\label{fig:A1664profs}
\end{centering}
\end{figure*}

\begin{figure}
\includegraphics[width=7.5cm,angle=0]{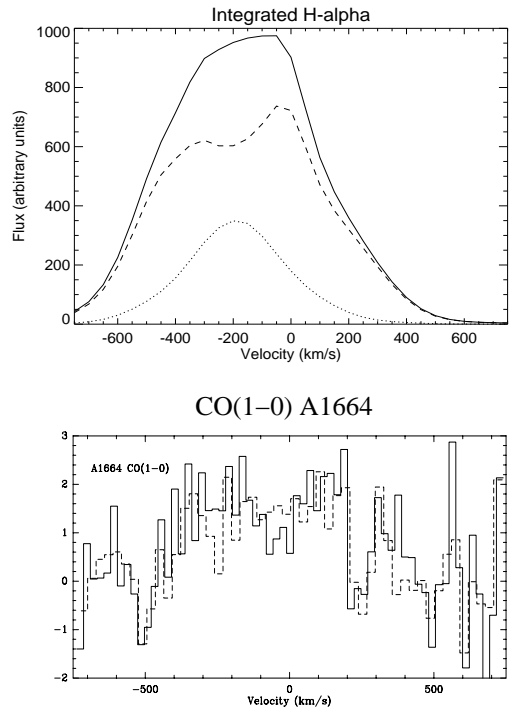}
\caption{The H$\alpha$ and CO(1-0) emission line profiles integrated over comparable areas of A1664. The CO(1-0) is derived from the IRAM 30m observation by Edge~(2001) with a 23.5\arcsec~beam size (the solid and dashed lines show measurements with different correlators). The H$\alpha$ 
profile was constructed by summing over the whole source and using the multi-component gaussian fits to the H$\alpha$+[NII] complex-- the solid shows all the emission, whilst the dotted and dashed line show the contributions from fibres requiring one and two velocity components, respectively. Velocities are with respect to the nominal cluster redshift of $z=0.1276$.}
\label{fig:A1664COHa}
\end{figure}

\begin{figure*}
\begin{centering}
\includegraphics[width=5.5cm,angle=90]{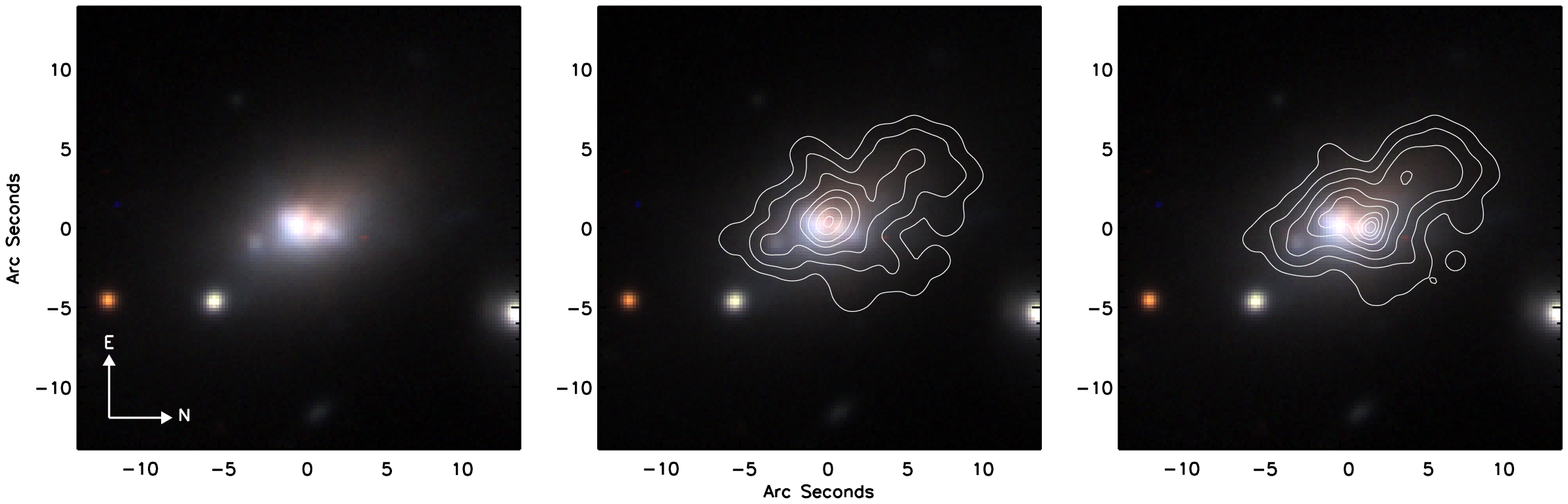}
\caption{Left: true-colour image of A1664 derived from archival VLT-FORS1 V, R and I imaging. Centre: overlaid with contours of the red H$\alpha$ velocity component of Fig.~\ref{fig:A1664FLUX}. Right: overlaid with contours of the blue H$\alpha$ velocity component.}
\label{fig:A1664FORS1}
\end{centering}
\end{figure*}

\section{A1835}
At a redshift of $z=0.2523$, A1835 is the most distant cluster presented in this study. It is also our most luminous H$\alpha$ emitter, with an extinction-corrected luminosity of $3 \times 10^{42}$\ergps~and a visible star formation rate of $\sim 100$\Msunpyr~(Crawford et al.~1999; Allen~1995). The {\em Spitzer} MIPS photometry of Egami et al.~(2006) reveal far infrared thermal dust emission with $L_{\rm{IR}} = 7 \times 10^{11}$\Lsun, plausibly powered by star formation. In the X-rays, the cluster is the most luminous in the ROSAT BCS and subsequent {\em Chandra} data are consistent with a young ($6 \times 10^{8}$\yr) cooling flow operating within $r=30$\kpc~(Schmidt, Allen \& Fabian~2001).

\begin{figure*}
\begin{centering}
\includegraphics[width=12cm,angle=0]{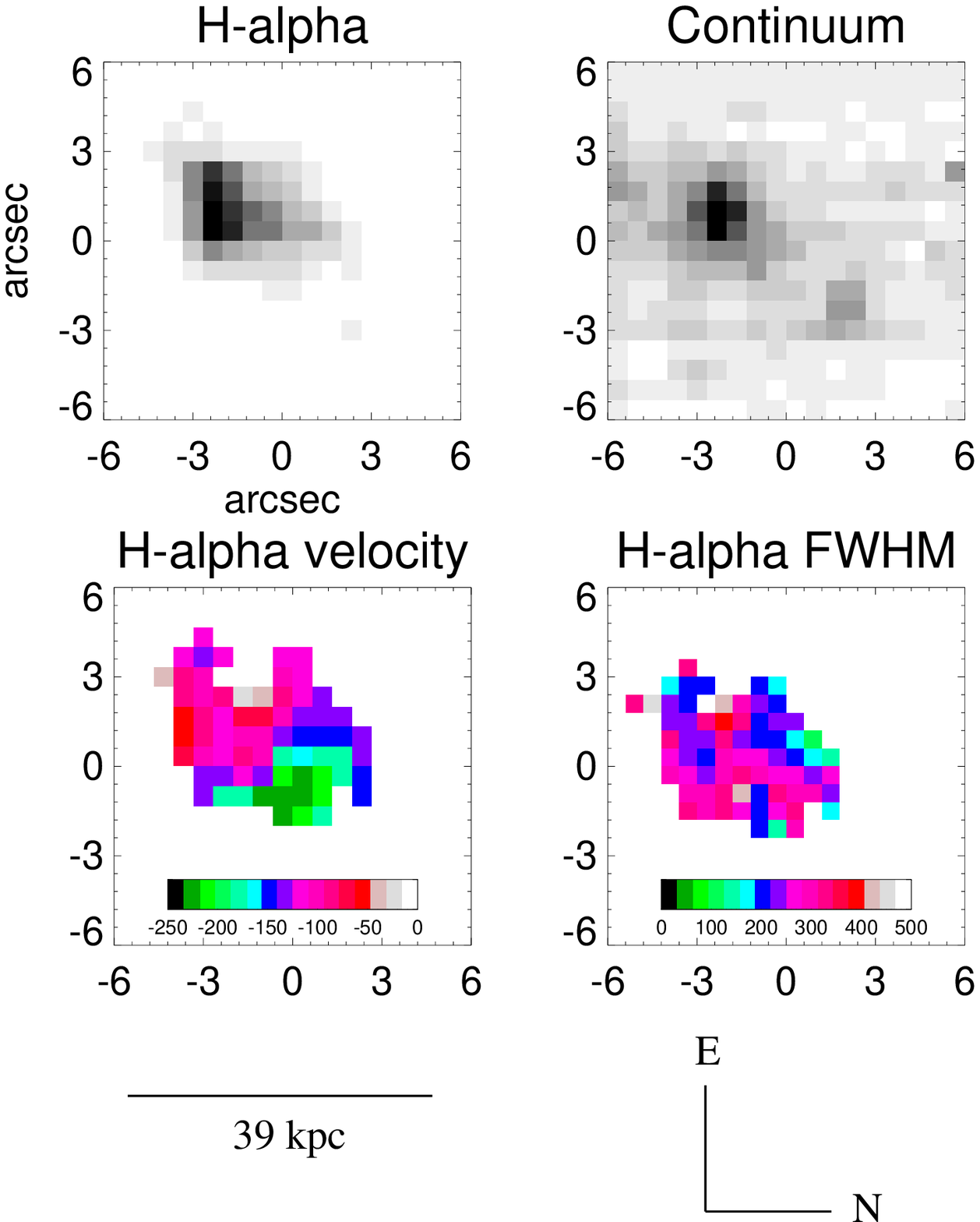}
\caption{Continuum (6600--6800\AA, observed frame), H$\alpha$ intensity, velocity field and FWHM for A1835, as derived from single component fits to the H$\alpha$+[NII] complex. The velocities are in units of \kmps~and relative to the nominal cluster redshift of $z=0.2523$. As explained in the text, only part of the field of view of one of the two dithered pointings was used.}
\label{fig:A1835FLUX}
\end{centering}
\end{figure*}

In comparison with A1664 the kinematics are simple and relatively quiescent, requiring a single velocity component at each position. The H$\alpha$ map and velocity field are shown in Fig.~\ref{fig:A1835FLUX}. The H$\alpha$ emission has a maximum extent of 8\arcsec~(31\kpc) and is elongated in a NW-SE direction; the reconstructed continuum image shows a small companion galaxy 5\arcsec~(20\kpc) north-west of the CCG. The emission velocity exhibits a shear of $\sim 250$\kmps~along the same axis, possible due to rotation, with linewidths of $\simeq 250$\kmps~FWHM.

\begin{figure}
\includegraphics[width=9.5cm,angle=0]{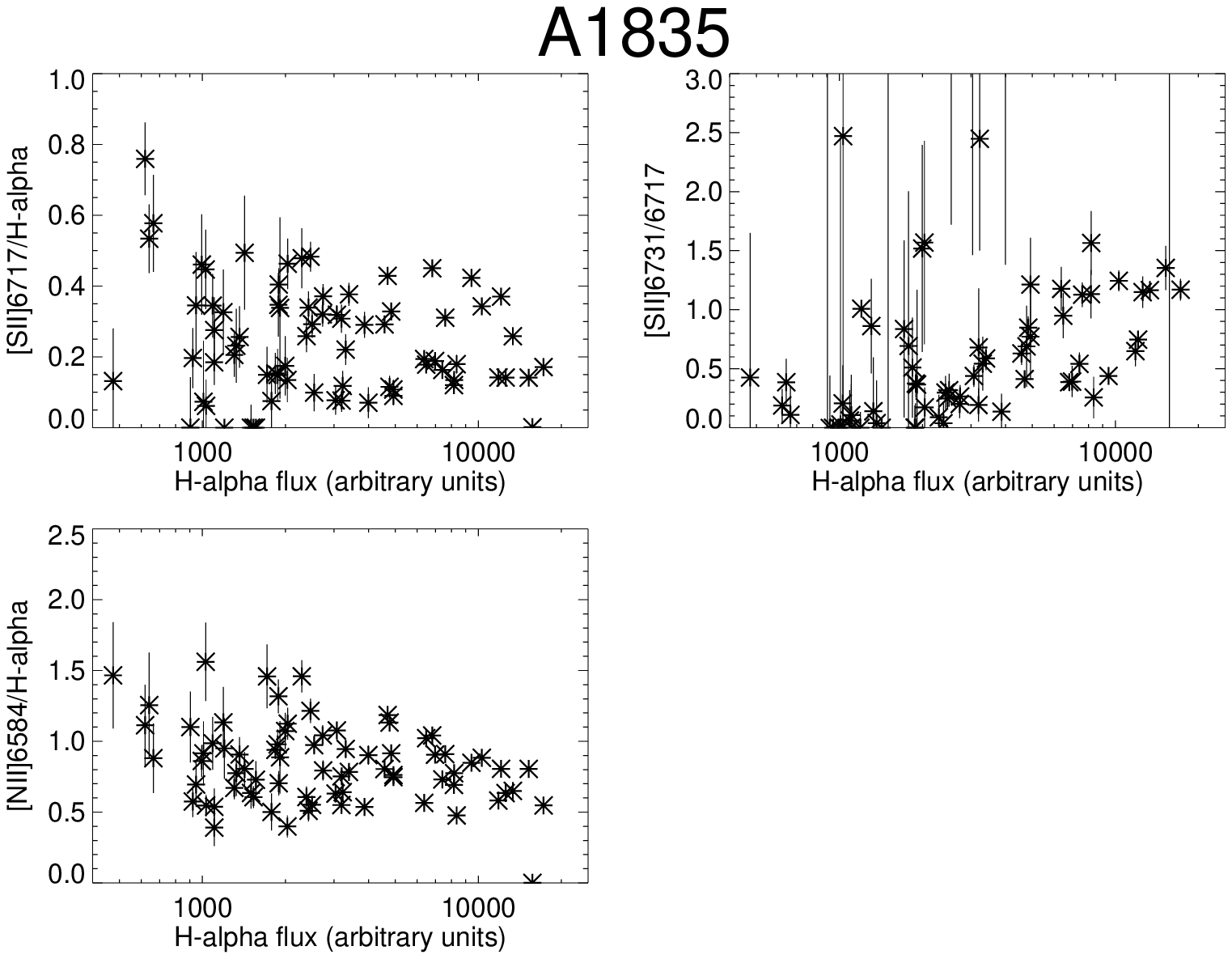}
\caption{Ionization diagnostic line ratios plotted as a function of H$\alpha$ flux for lenslets in A1835.}
\label{fig:A1835ions}
\end{figure}

As for A1664, the [NII]+H$\alpha$ kinematic fits were used to search for associated emission in [SII]$\lambda\lambda6717,6731$. Unfortunately, the emission in [OI]$\lambda\lambda6300,6363$ is badly affected by night sky emission line features and was not analysed.  Line ratios as a function of local H$\alpha$ flux are shown in Fig.~\ref{fig:A1835ions}, but the only firm conclusion that can be drawn is that the [NII]$\lambda6584$/H$\alpha$~is once again remarkably constant with luminosity and consistent with the integrated slit value measured by Crawford et al.~(1999). The [SII]$\lambda 6731/6717$ ratio shows marginal evidence for an increase with H$\alpha$ surface brightness, implying higher electron density as hinted at in A1664. 

A comparison with the CO emission is once again valuable. The IRAM 30m single dish observation of Edge~(2001) 
shows CO(1-0) from an inferred cool molecular mass of $1.8 \pm 0.2 \times 10^{11}$\Msun. The CO emission peaks at a velocity of $-100$\kmps~(relative to an assumed redshift of $z=0.2523$), identical to that of the H$\alpha$ emission peak. The CO linewidth of $227 \pm 38$\kmps~FWHM is again a close match to the H$\alpha$. Subsequent Owen Valley Millimetre Array interferometry of the CO(1-0) emission constrains the angular extent of the CO to $<9$\arcsec~(36~kpc). It thus appears that, as in A1664, the CO and H$\alpha$ emission share the same kinematics and morphology.

\section{A2204}
Abell 2204 lies at $z=0.1514$ and has an observed H$\alpha$ slit luminosity of $10^{42}$~\ergps~(uncorrected for reddening), although the inferred visible star formation rate is modest ($\sim 1$\Msunpyr~Crawford et al.~1999). The system is also a strong emitter of near-infrared H$_{\rm{2}}$ ro-vibrational emission (Edge et al.~2002), and the CO(1-0) emission from {\em IRAM} 30-m observations implies a molecular gas mass of $2.3 \pm 0.6 \times 10^{10}$\Msun~(Edge~2001). {\em Chandra} X-ray observations reveal that the cluster has a disturbed core morphology, suggestive of a recent merger, with cold fronts at radii of $\sim 28$~kpc and 54.5\kpc, and the central radio source is also disturbed with three components within 10 arcsec, aligned roughly N-S (Sanders, Fabian \& Taylor~2005).  

\begin{figure*}
\begin{centering}
\includegraphics[width=12.5cm,angle=0]{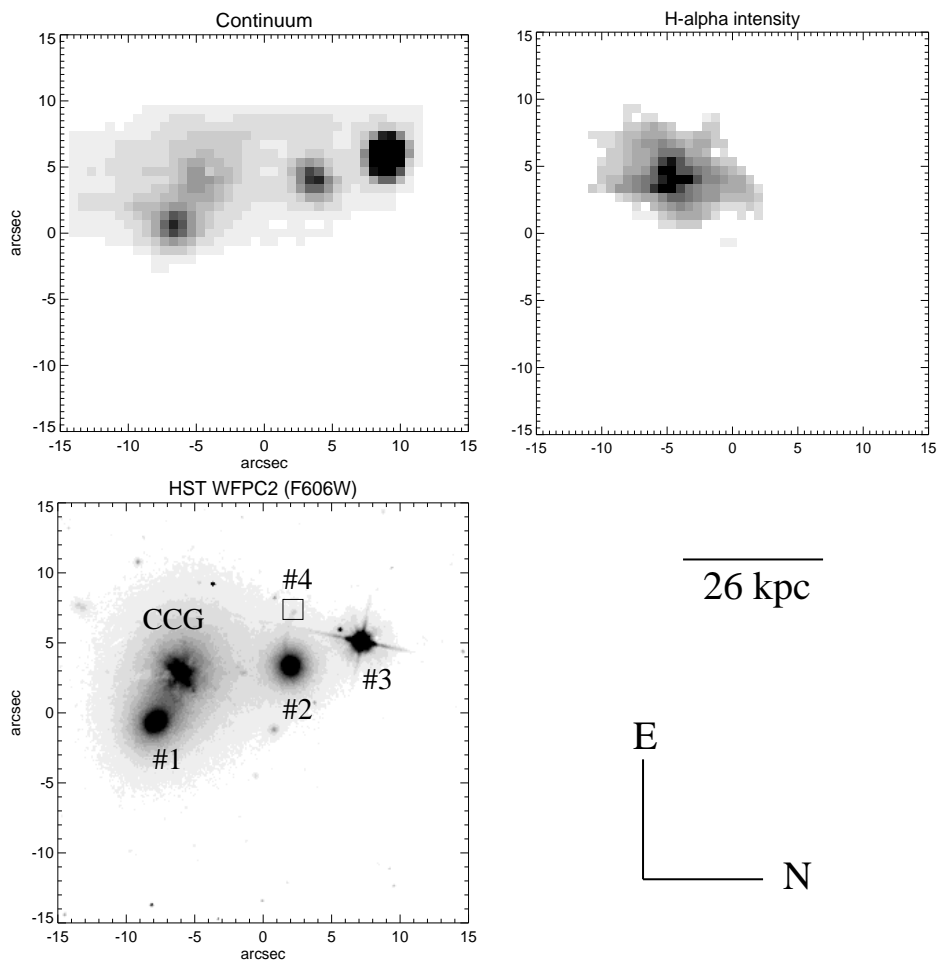}
\caption{H$\alpha$ intensity and 7000-7200\AA~(observed wavelength) continuum in A2204. The former is derived from a single component fit to the H$\alpha$+[NII] complex. For comparison, we also shown an HST WFPC2 F606W image of the same region. In the latter, we identify three separate continuum objects distinct from the central cluster galaxy (CCG) and a $z=1.06$ lensed background [OII] emitter (object \#4) (shown in a magnified portion of the HST frame in Fig.~\ref{fig:A2204gravlens}), all of which are discussed further in section 5.}
\label{fig:A2204FLUX}
\end{centering}
\end{figure*}

Results from the VIMOS data in H$\alpha$ and the continuum are shown in Fig.~\ref{fig:A2204FLUX}, along with an 
HST WFPC2 F606W (wide V-band) continuum image. The latter shows that the CCG is dusty and possibly interacting with galaxy \#1, which is 15\kpc~away in projection at a relative velocity of 250\kmps~(see Jenner~1974 and section 5.1). The H$\alpha$ emission also has an irregular morphology with a number of filaments emanating from the nuclear regions, possibly coincident with the dust obscuration. The H$\alpha$ velocity field has a tri-partite structure with three distinct clumps at velocities of $\simeq$--200, --50 and 100\kmps~(relative to $z=0.1514$). The velocity width of the H$\alpha$ is in most locations $\simeq 200$\kmps~FWHM, except in the nuclear region where it rises to $\simeq 800$\kmps. This increased velocity width is unlikely to be due to an AGN because the emission line ratios are uncharacteristic of AGN and show no spatial variation across the source. The {\em Chandra} X-ray data show no evidence for a nuclear point source. A similarly high velocity width in a spur to the north-east of the nucleus may be due to interaction between the emission line clouds and the radio source, whose core structure extends 5~arcsec from the nucleus in this direction. The high H$\alpha$ line width may explain the apparent faintness of the CO(1-0) line (Edge~2001). VLT-ISAAC long-slit spectroscopy by Jaffe, Bremer \& Baker~(2005) of A2204 shows that the 
Pa$\alpha$ and 1-0~S(1) trace each other very closely, spatially and kinematically. Narrow-band H$\alpha$ imaging by the same authors shows that the line emission drops off quite sharply, being characterised by a gaussian of FWHM $\simeq 30$\kpc. Other clusters, in contrast, show a power-law emission extending out to radii of 40\kpc~and beyond.

As remarked above, the line ratios exhibit no spatial variations so are plotted as a function of local H$\alpha$ flux in Fig.~\ref{fig:A2204ions}. The constancy of the [NII]$\lambda 6584$/H$\alpha$ and [OI]$\lambda 6300$/H$\alpha$ is striking. As in the other targets, there is marginal evidence for an increase in [SII]$\lambda6717$/H$\alpha$ and [SII]$\lambda6731$/6717 with increasing H$\alpha$ surface brightness.

\begin{figure*}
\begin{centering}
\includegraphics[width=14.5cm,angle=0]{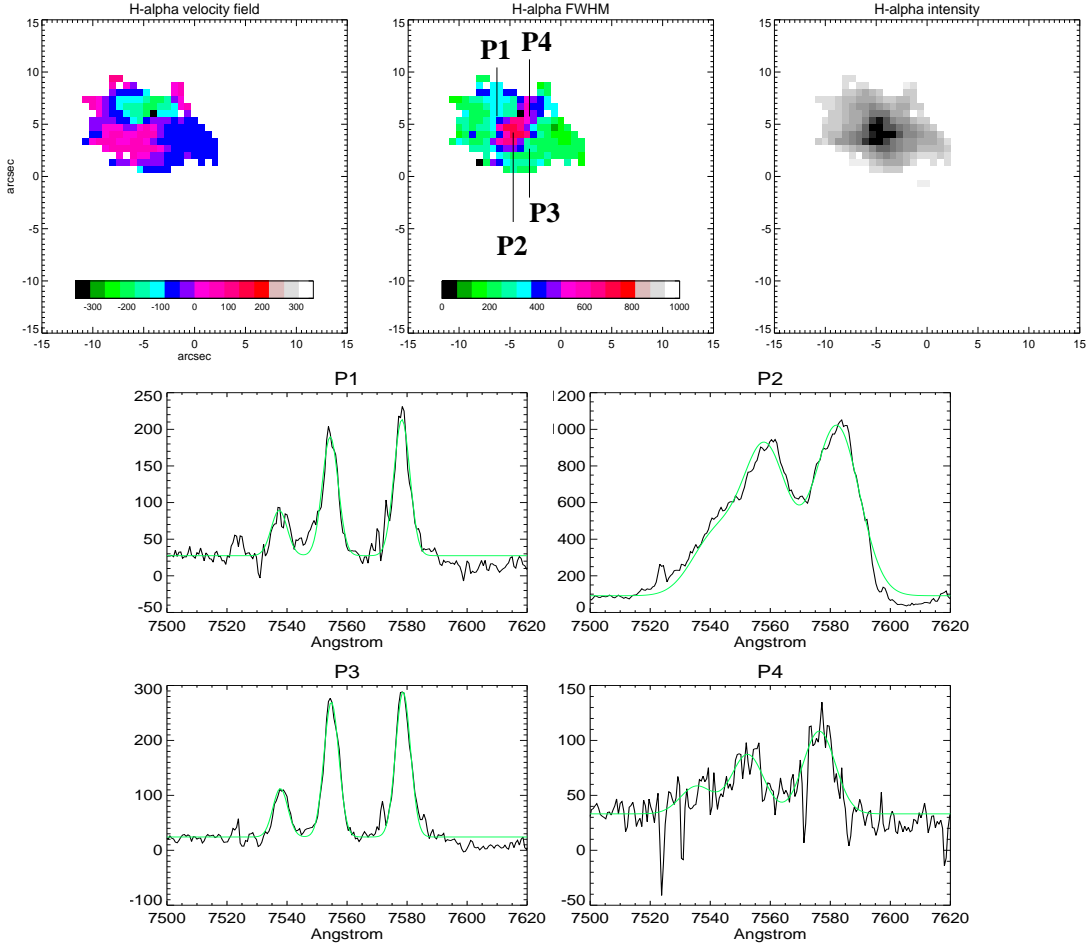}
\caption{Velocity field and FWHM (in \kmps) derived from fits to the H$\alpha$+[NII] complex in A2204; the H$\alpha$ intensity map of Fig.~\ref{fig:A2204FLUX} is reproduced here for ease of comparison. The zero-point of the velocity field is taken to be the nominal cluster redshift of $z=0.1514$. Line profiles for some individual fibres are also shown.}
\label{fig:A2204profs}
\end{centering}
\end{figure*}

\begin{figure}
\includegraphics[width=9.5cm,angle=0]{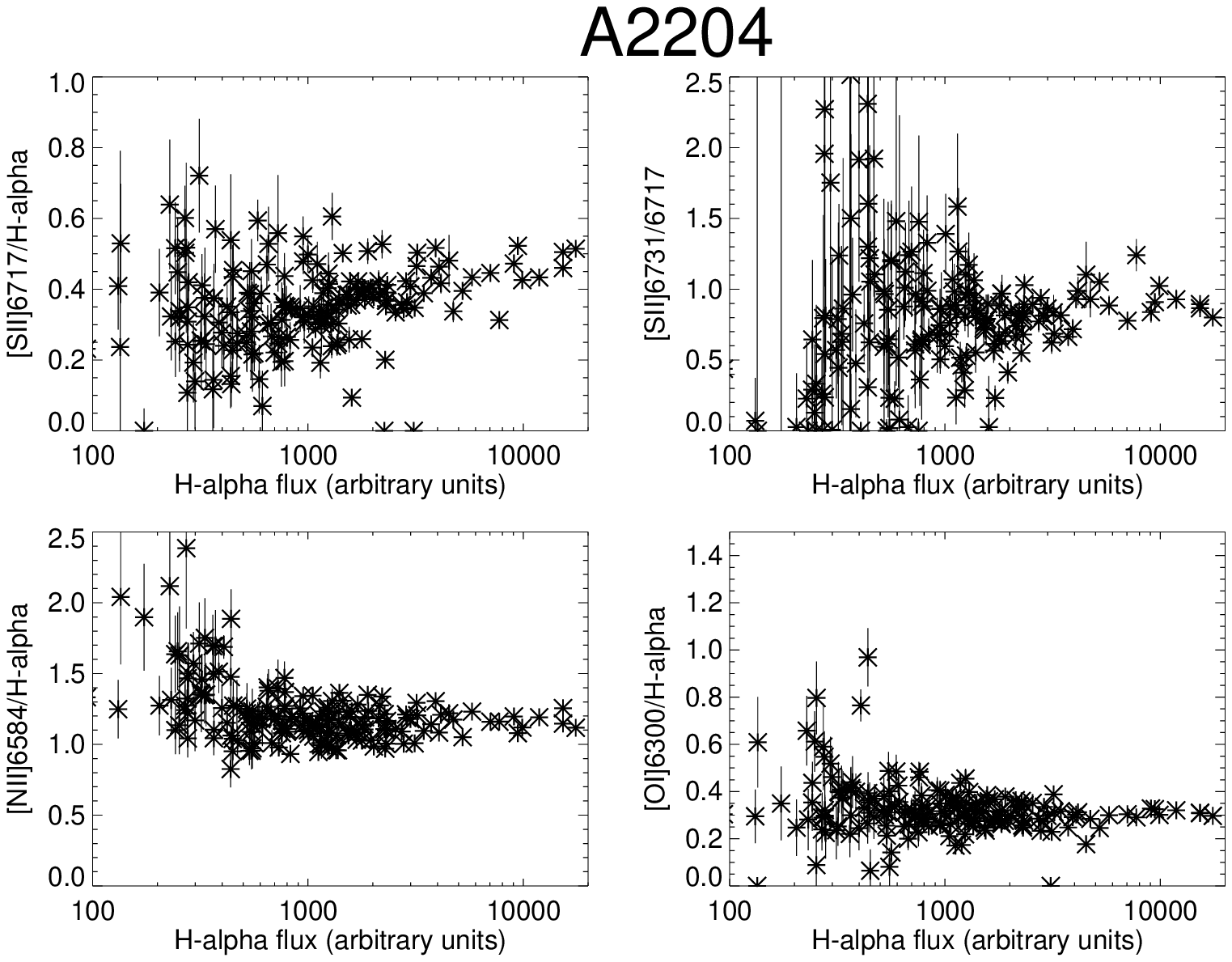}
\caption{Ionization diagnostic line ratios plotted as a function of H$\alpha$ flux for lenslets in A2204.}
\label{fig:A2204ions}
\end{figure}

\subsection{Continuum sources in the field of A2204}
\begin{figure}
\includegraphics[width=7.5cm,angle=0]{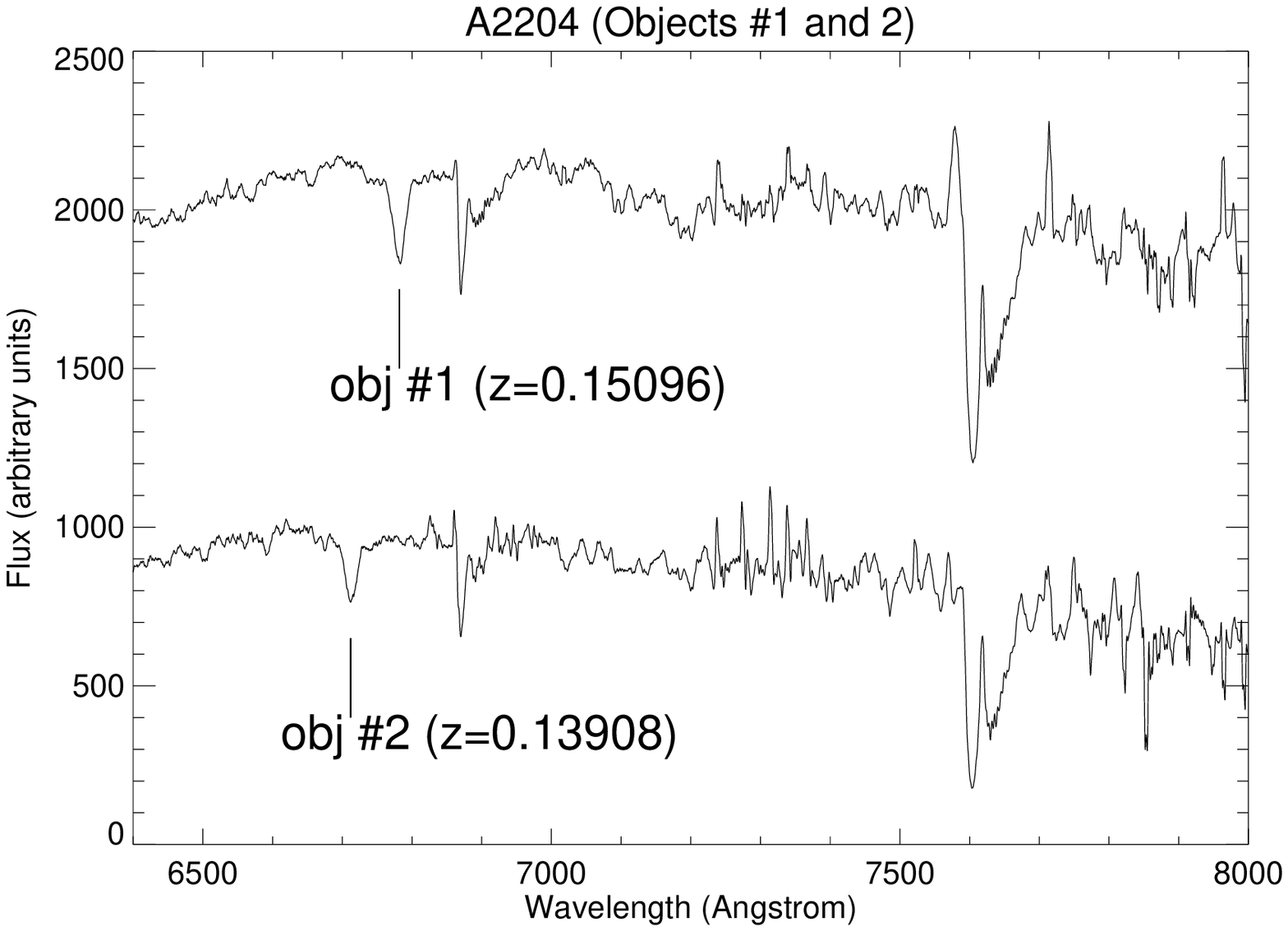}
\includegraphics[width=7.5cm,angle=0]{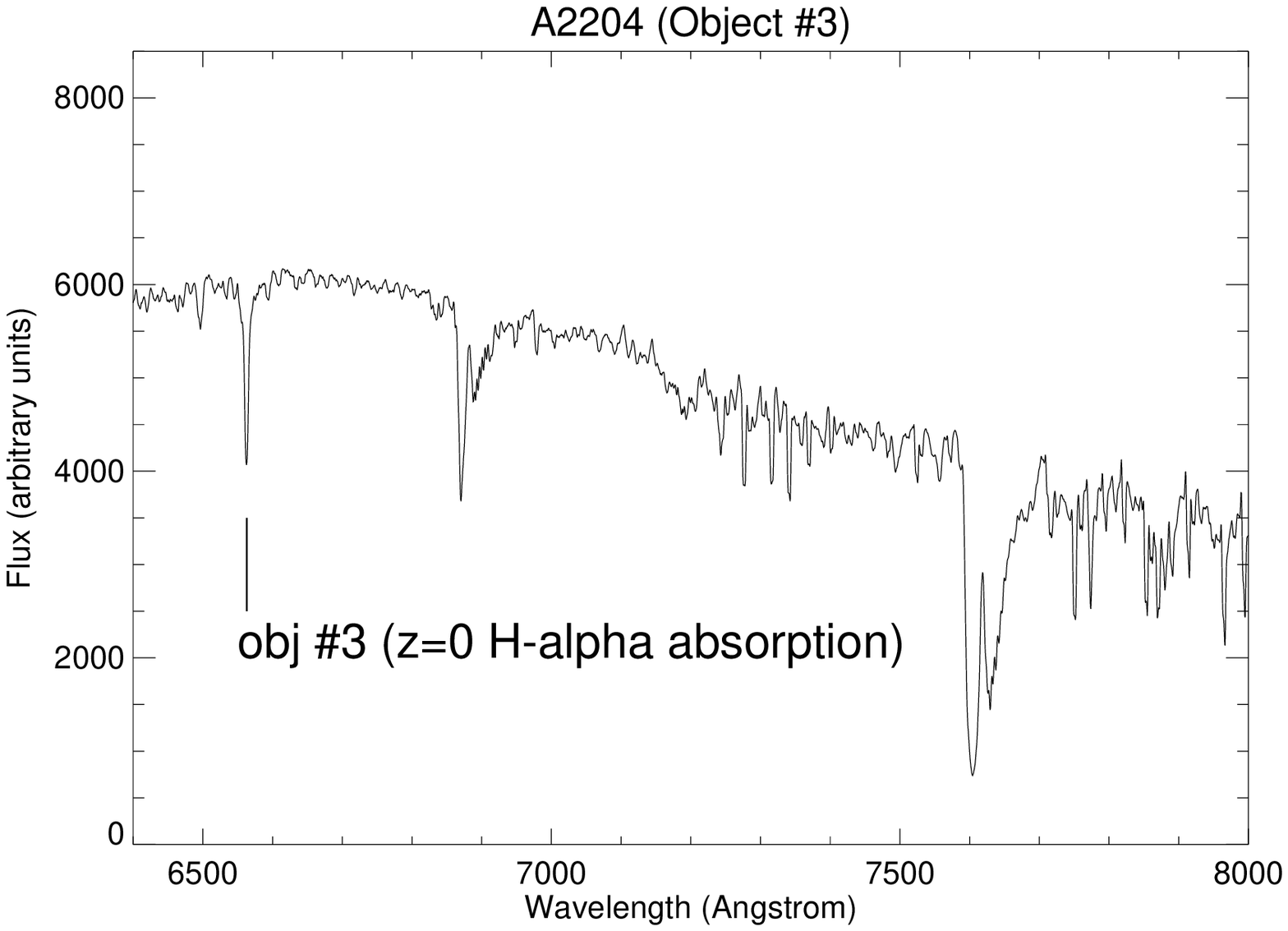}
\caption{Spectra of continuum objects \#1--3 in the VIMOS field of view of A2204 (see Fig.~\ref{fig:A2204FLUX}). Redshifts were 
determined from the Na D$\lambda$5892.5 absorption blend (\#1 and 2) and confirm their cluster membership. Object \#3 is a star. These spectra have not been flux calibrated in either an absolute or relative sense. The features between 6860--6880\AA~and 7600--7700\AA~are telluric absorption.}
\label{fig:A2204contOBJ}
\end{figure}

In Fig.~\ref{fig:A2204contOBJ} we show spectra of the three continuum sources visible in the VIMOS field of view (Fig.~\ref{fig:A2204FLUX}) alongside the CCG. The HST WFPC2 image of the same field shows that objects \# 1 and 2 are galaxies and object \#3 stellar. From the Na D$\lambda$~5892.5 absorption blend we determined redshifts of z=0.15096 (\#1) and z=0.13908 (\#2); the presence of z=0 H$\alpha$ absorption confirms the stellarity of object \#3. Galaxy \#1 is thus close in projection ($\simeq 15$\kpc) and velocity ($\simeq 100$\kmps) to the CCG of A2204. It is thus possible that the two are interacting. Galaxy \#2 is twice as far away in projection and offset by --3200\kmps~in velocity.

\begin{figure}
\includegraphics[width=6.5cm,angle=0]{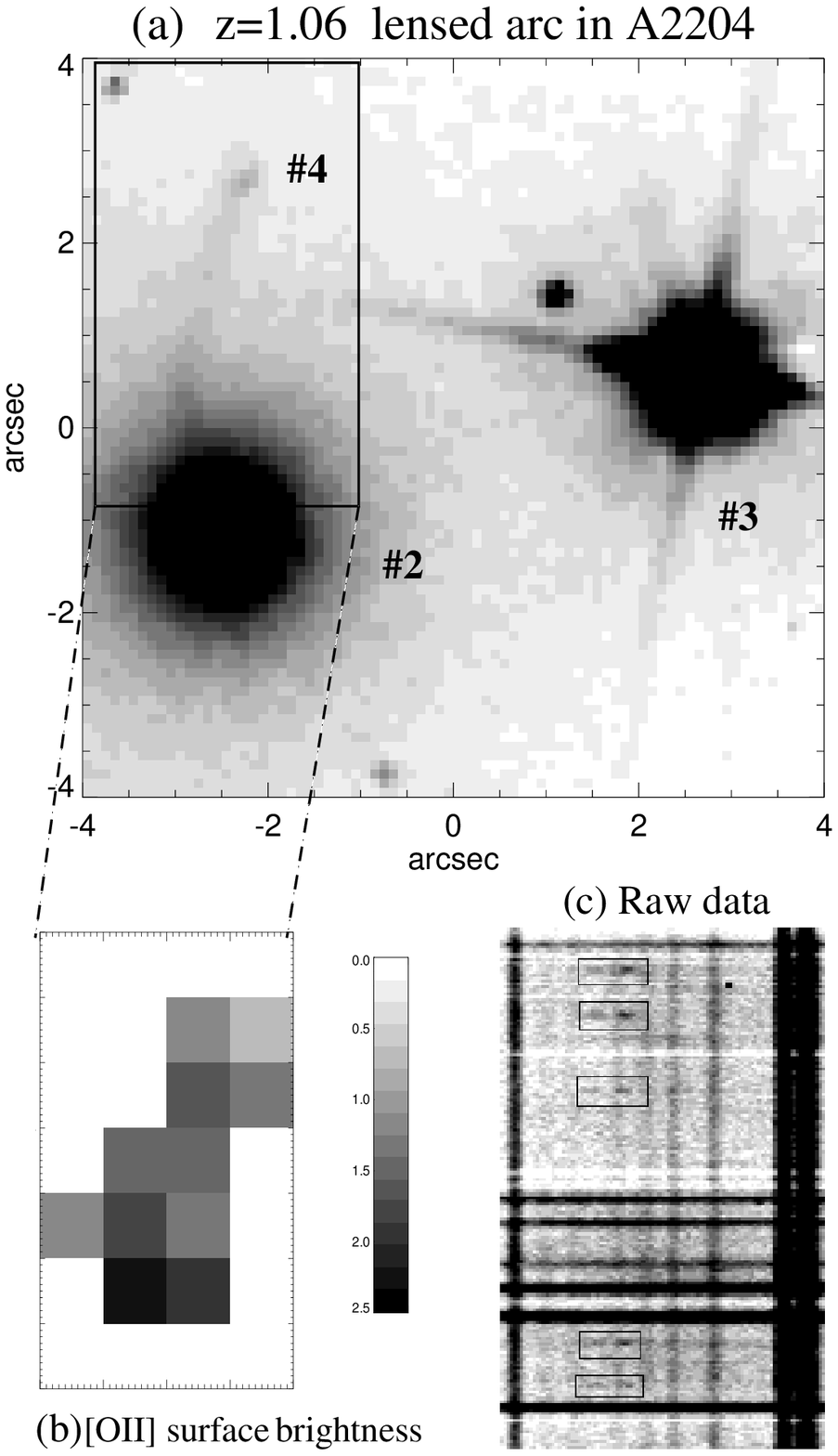}
\caption{The $z=1.0604$ gravitationally-lensed background galaxy in the field of A2204. (a) Part of the HST WFPC2 F606W frame showing the arc in the continuum; objects 2--4 are labelled as in Fig.~\ref{fig:A2204FLUX}. (b) Reconstructed image in the [OII]$\lambda \lambda$3726,3729 doublet using the VIMOS data; the surface brightness is in units of $10^{-16}$~erg~cm$^{-2}$~s$^{-1}$~arcsec$^{-2}$. (c) A portion of the wavelength-calibrated 2D VIMOS spectral frame of A2204 in which the arc was discovered; boxes higlight the [OII]$\lambda \lambda$3726,3729 emission.}
\label{fig:A2204gravlens}
\end{figure}

\subsection{The gravitationally-lensed background galaxy}
The {\em HST} image of the cluster in Fig.~\ref{fig:A2204FLUX} exhibits a gravitationally lensed arc which terminates on a bright knot of emission which we refer to as object \#4. The VIMOS data show [OII]$\lambda \lambda 3726,3729$ emission in 11 fibres around this position which, with a redshift of $z=1.0604$, we identify as the lensed background galaxy. Fig.~\ref{fig:A2204gravlens} shows the portion of the calibrated 2D spectral frame in which this emission was discovered, together with the reconstructed emission line image. The line is just at the edge of the 7600--7700\AA~telluric absorption, for which we correct with the aid of the stellar continuum spectrum of object \#3. Line emission is present along the full length of the arc as defined by the {\em HST} image. The total [OII] flux from it is estimated to be $7.2 \times 10^{-16}$\ergpcmsqps, a figure obtained by comparing our H$\alpha$ measurments of the CCG with the slit flux in Crawford et al.~(1999). The velocity of the emission varies by less than $\pm 20$\kmps, and the line width is the range 100--150\kmps~FWHM. This lensed arc would be suitable for a follow-up with a longer IFU observation at higher spatial resolution, possibly with adaptive optics due to the proximity of the moderately bright star \#3 (R=15.3; B=16.5 mag).

\section{Zw~8193}
\begin{figure*}
\begin{centering}
\includegraphics[width=10.0cm,angle=90]{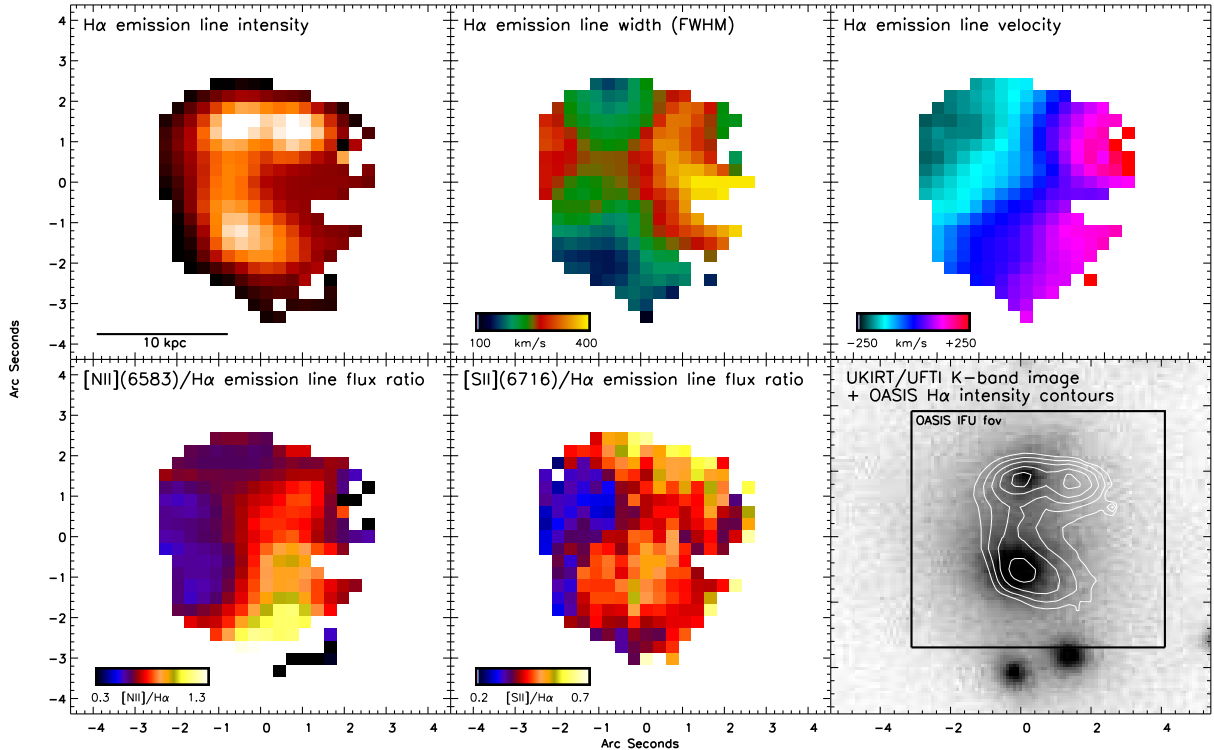}
\caption{The H$\alpha$+[NII] and [SII] line emission in Zw~8193 derived from the WHT/OASIS observations, compared with a K-band continuum image. 
North is up, east is to the left.}
\label{fig:Zw8193}
\end{centering}
\end{figure*}

The core of Zw~8193 is complex in both space and velocity and initial optical spectroscopy of this cluster yielded discrepant redshifts for the emission ($z=0.1829 \pm 0.0001$) and absorption line ($z=0.1725 \pm 0.0001$) components (Allen et al.~1992). Fig.~\ref{fig:Zw8193} shows that there are three galaxies within 2--3 arcsec of the CCG; UKIRT long-slit spectroscopy revealed Pa$\alpha$ emission in and between the CCG and its neighbour to the north, with an apparent velocity difference of $400-500$\kmps~(Edge et al.~2002). The benefit of higher spatial and spectral resolution OASIS IFU data now reveals the full complexity of the system. The H$\alpha$ emission consists of three distinct `blobs', one centred on the CCG, one on a galaxy to the north and one slightly offset from the latter to the west. The velocity ranges over $\pm 250$\kmps~(relative to $z=0.1825$) and reconstructing the placement of the 1.2\arcsec-wide N-S slit used for the UKIRT observations shows the consistency of the Pa$\alpha$ and H$\alpha$ velocity fields. The [NII]/H$\alpha$ ratio varies from 0.9 on the CCG to 0.7 at the position of the northern blobs. The ratio [SII]$\lambda6717$/H$\alpha \simeq 0.5$ and is also relatively constant across the emission. The [SII]$\lambda6731$ line is badly affected by a night-sky emission line. The extinction-corrected H$\alpha$ slit luminosity quoted by Crawford et al.~(1999) is $3.1 \times 10^{42}$\ergps~(converted to the cosmology used here).

The CO(1-0) observations of Zw~8193 have yielded seemingly inconsistent results (Edge~2001). One observation 
revealed a broad plateau of emission from $v=-100$ to $v=+500$\kmps~but a subsequent observation failed to 
confirm it. As remarked by Edge~(2001), the presence of a flat-spectrum radio source in Zw~8193 complicates 
the baseline subtraction. The quoted upper limit on the molecular gas mass is $4.3 \times 10^{10}$\Msun. 

As in A1664, and to a lesser extent A2204, the H$\alpha$ emission morphology and kinematics appear to be influenced by the close passage 
of a small cluster galaxy into the core.

\section{SUMMARY OF OBSERVATIONAL RESULTS} 
Despite the diversity and complexity of these individual clusters and the small sample size, a 
number of generic features are apparent. They are:

\noindent
{\em $\bullet$ A possible association between disturbed H$\alpha$ emission and secondary galaxies within 10--20\kpc}. 
In A1664 it appears plausible on both morphological and kinematic grounds that the H$\alpha$ filament and complex 
nuclear structure in the CCG have been produced by the infall of a small cluster galaxy. This galaxy is now visible at the extremity 
of the filament. Since the mass of H$\alpha$-emitting ionized gas is relatively small (a few $10^{7}$\Msun, as inferred from the H$\alpha$ luminosity and the measured electron density), it is possible that a significant fraction of it may have been stripped from the infalling galaxy. Our 
preferred interpretation, discussed in section 8, is that gas distribution has been disturbed by the infalling galaxy. In Zw8193 there is 
extended H$\alpha$ emission associated with a galaxy which lies within 6\kpc~(projected) of the CCG. In A2204, the irregular H$\alpha$ kinematics 
may result from a recent interaction with a nearby galaxy (\#1 in Fig.~\ref{fig:A2204FLUX}), or disturbance from the irregular radio source. In 
A1835, the H$\alpha$ emission is elongated in the direction of a small companion galaxy 20\kpc~away. Another well-studied CCG showing strong 
evidence for the influence of a nearby secondary galaxy on the H$\alpha$ emission is RXJ 0820.9+0752 at $z=0.110$ (Bayer-Kim et al.~2002).

\noindent
{\em $\bullet$ Uniformity of the ionization state.} 
Despite the morphological and kinematic complexity of the H$\alpha$ emission, the ionisation state of the gas is 
surprisingly uniform, as a function of position, H$\alpha$ surface brightness and between clusters. The [NII]$\lambda6584$/H$\alpha$ and [OI]$\lambda6300$/H$\alpha$ ratios are particularly tight, but there may be hints of a weak increase of 
[SII]$\lambda 6717$/H$\alpha$ and [SII]$\lambda 6731/6717$ with H$\alpha$ surface brightness. These results suggest
that the same mechanism(s) excite all the optical line emission in these systems, and that they are quite independent 
of the processes which disturb the gas morphologically and kinematically on the kiloparsec scales we observe, i.e. an 
infalling galaxy or a weak radio source may stir up the gas, but they seem not to impact on its 
ionization state. For this reason, it seems unlikely that large-scale shocks are involved. 

The line ratios we measure are naturally consistent with the integrated slit values found by Crawford et al.~(1999) 
and characteristic of high H$\alpha$ luminosity ($> 10^{41}$\ergps) cooling flow CCGs. The latter show strong 
AGN-like [NII]$\lambda6584$/H$\alpha$ and [OI]$\lambda6300$/H$\alpha$, but weak [SII]$\lambda 6717$/H$\alpha$ 
and [OIII]$\lambda 5007$/H$\beta$ similar to starbursts. As mentioned in section 1, it was shown by Crawford et al.~(1999) 
that the observed H$\alpha$ luminosities of such systems can be powered by photoionisation by the O-star populations which explain
their excess blue light. Typically $10^{5}-10^{6}$ 0 stars and larger populations of B, A and F stars are required in excess of the continuum
expected from a cluster elliptical. However, the ro-vibrational H$_{\rm{2}}$:Pa$\alpha$ line ratios of CCGs are much higher than 
typical starburst galaxies (Jaffe, Bremer \& van der Werf 2001; Edge et al.~2002). This may require a distinct 
population of much denser clouds (Wilman et al.~2002) or excitation by a spectrum harder than that of O-type stars 
(Jaffe, Bremer \& Baker 2005). This may also explain why the optical emission line ratios differ from those of 
starburst galaxies. For the ensemble of line-emitting CCGs, Crawford et al.~(1999) discovered a continuous trend 
of decreasing [NII]$\lambda6584$/H$\alpha$ with increasing H$\alpha$ luminosity, and similar weaker trends in 
[OIII]$\lambda 5007$/H$\beta$ and [SII]$\lambda6717$/H$\alpha$. Our results show that similar trends do not extend to 
{\em local} H$\alpha$ surface brightness within the most luminous systems.  

Sparks et al.~(1989) proposed that cool gas (e.g. from a galaxy merger) falling into the core
of a galaxy cluster could be energised by thermal conduction from the X-ray-emitting ICM. This would lead
to optical line emission from the cooler gas and a local enhancement in X-ray emission. Such a model was used to 
account for the close correspondence between the X-ray and H$\alpha$ filaments in the Virgo and Perseus clusters 
(Sparks et al.~2004). It is not known whether this mechanism can reproduce the observed emission line ratios.

\noindent
{\em $\bullet$ CO and H$\alpha$ trace the same gas, at least kinematically and probably also spatially.}
Two of the four clusters, A1664 and A1835, have robust detections of CO(1-0) emission in the survey of Edge~(2001). 
In both these clusters the CO emission profile is in very good agreement with the H$\alpha$ kinematics, in terms of its centroid 
velocity and line width, strongly suggesting that the CO and H$\alpha$ emission arise from the same ensemble of clouds. 
Given this strong CO:H$\alpha$ link, the weak (A2204) or ambiguous (Zw8193) CO detections in Edge~(2001) most likely 
stem from a high H$\alpha$ velocity width or zero-point offset (relative to the CCG redshift then assumed), respectively.  
In the aforementionned RXJ 0820.9+0752 at $z=0.110$ (Bayer-Kim et al.~2002) the CO and extended H$\alpha$ kinematics are also well-matched. 
This is in line with the correlation between H$\alpha$ luminosity and molecular gas mass discovered by Edge~(2001). In contrast, the ro-vibrational 
H$_{\rm{2}}$ emission and kinematics appear to be less strongly coupled to the H$\alpha$. In A1664, the H$_{\rm{2}}$ has a much lower velocity 
width than Pa$\alpha$ and CO, although in A2204 the match is much closer. RXJ 0820.9+0752 has weak or non-existent H$_{\rm{2}}$ emission in the 
Edge et al.~(2002) survey, despite its strong extended CO and H$\alpha$ emission.

We note, however, that Jaffe, Baker \& Bremer~(2005) discovered that the Pa$\alpha$ and hot \H2~trace each other spatially and kinematically out 
to radii beyond 20\kpc, but A2204 is the only cluster common to our two samples.

\section{INTERPRETATION: PHYSICAL PROCESSES IN THE DENSE MOLECULAR GAS RESERVOIR}
We now interpret the above findings, taking as our starting point the CO observations. As reviewed in section 1, 
these have revealed substantial masses of cool molecular gas ($10^{9-11.5}$\Msun, at 20--40\K) in the cores of cooling 
flows, with a mass which correlates with the H$\alpha$ luminosity (Edge~2001). Subsequent interferometry has shown that 
the CO emission is confined to scales $<20$\kpc~(Edge \& Frayer~2003). The lack of double-peaked CO line profiles 
suggests that the molecular cloud distribution is quasi-spherical, rather than disk-like. Such a dense, compact 
molecular gas reservoir will most likely be disturbed by the infall of a small galaxy into the cluster core, as we 
believe is happening most clearly in A1664, or by the onset of a nuclear radio source. We consider these possibilities in more detail 
below. First we consider some physical processes in the unperturbed molecular gas reservoir and their observable 
consequences.

It is plausible that the CO and H$\alpha$ emission arise within the same clouds, because they appear to 
share the same kinematics and are closely correlated in luminosity. The ionised surfaces of such clouds are likely to 
be in pressure equilibrium with the hot X-ray gas for which $n \sim 0.1$\pcm, $T \sim 10^{7}$\K; for optical emission line
regions $n  \sim 100$\pcm, $T \sim 10^{4}$\K, such that $nT$ is the same for both phases. Pressure equality may also 
extend to the cool molecular gas if $n \sim 10^{5}$\pcm, although it is possible that 
the cloud cores are self-gravitating. The clouds may be warmed/ionised by a combination of young stars and X-ray 
emission/conduction from the intracluster medium, but the precise details do not concern us here. The 
rovibrational H$_{\rm{2}}$ emission may be a distinct, transiently-heated high-pressure component, since line 
ratio analysis points to thermal excitation in dense gas with $n > 10^{5}$\pcm~and $T \sim 2000$\K~(Jaffe, Bremer, van 
der Werf~2001; Wilman et al.~2002). Wilman et al.~(2002) proposed that shocks between a population of low-density 
($\sim 200$\pcm) H$\alpha$-emitting clouds could be used to transiently create such high-density molecular gas, although 
an external stellar radiation field was still used to excite the observed emission. We now modify this model and 
investigate collisions amongst the denser CO clouds.

\subsection{Cloud-cloud collisions and external perturbations}
Consider that a mass $M$ of CO-emitting molecular gas is confined within a radius $R_{\rm{0}}$ of the CCG, 
distributed within an ensemble of identical clouds of radius $r$, mass $m_{\rm{c}}$, space density $\cal{N(R)}$ and 
hydrogen number density $n$. Interferometry shows that the CO is centrally concentrated so we take 
$\cal{N(R)} \propto R$$^{-1}$, as in the inner part of an NFW profile (Navarro, Frenk \& White~1997). For a typical relative cloud 
velocity of $v$, the rate of cloud collisions per unit volume is $\cal{N(R)}$$^{2}$ $\sigma v$, where $\sigma$ is the collision 
cross-section, which we approximate with the geometrical cross-section $\sigma = \pi r^{2}$. It then follows that the rate at which mass
is processed through such collisions is:

\begin{equation}
\dot{M}_{COLL} = \frac{3 M^{2} v}{2 \pi^{2} R_{\rm{0}}^{3} n m_{\rm{p}} r}.
\end{equation}

For $v=150$\kmps, $R_{\rm{0}} = 10$\kpc, $r=1$\pc~and $n=10^{5}$\pcm, eqn.~(1) evaluates to
$\dot{M}_{COLL} \sim 1-100$\Msunpyr~for $M=10^{10-11}$\Msun~(m$_{\rm{p}}$ is the proton mass). Such supersonic 
collisions will shock the clouds to temperatures of several $10^{5}$\K. For the assumed densities the cooling time 
from these temperatures is very short, $\sim 1$~month, so the gas will quickly return to the cool phase, i.e. collisions 
will merely recycle the cool gas and not deplete it. It is, however, possible the shock compression will act as a 
trigger for star formation. They may also create the dense molecular gas and a locally strong, hard (e.g. black body at 
$T \geq 10^{5}$\K) radiation field required for the production of the rovibrational H$_{\rm{2}}$ emission. The EUV 
luminosity produced by such shocks is $L_{COLL} = \frac{1}{2} \dot{M}_{COLL} v^{2}$, or:

\begin{equation}
L_{COLL} = \frac{3 M^{2} v^{3}}{4 \pi^{2} R_{\rm{0}}^{3} n m_{\rm{p}} r}.
\end{equation}

For the same parameters as before, this amounts to $L_{COLL} = 7 \times 10^{39-41}$\ergps. Although this is several orders of magnitude below
the inferred luminosity of the young stellar component in these systems, the cubic dependence on $v$ and $R_{\rm{0}}$ and uncertainty over 
the actual values of $r$ and $n$ means that such shocks could make a non-negligible contribution to the excitation of the line emission in CCGs, 
at least locally if the gas clouds are not uniformly distributed. We note, however, that limits on the strength of the [OIII]$\lambda 4363$ emission 
line imply that shocks are unlikely to play a strong role in the production of the optical line emission in H$\alpha$-luminous CCGs 
(e.g. Voit \& Donahue 1997). The mechanism may, however, be a relatively more important source of excitation in lower-H$\alpha$ luminosity 
systems. Such shocks will steadily drain energy from the molecular gas reservoir, forcing the distribution to smaller radii and thus increasing 
the collision rate still further. 

We now consider the consequences of disturbing this reservoir with the infall of a small galaxy, as appears to 
be happening most dramatically in A1664. From the constant excitation state of the ionised gas, it seems that this does not 
cause any large scale shocks or cloud destruction. Most likely, the cloud distribution will be dynamically perturbed by the infalling galaxy, 
producing the observed streams and filaments. Such interactions are likely to enhance the star-formation rate, either by increasing the 
cloud-cloud collision rate (as above), or via some other means. With reference to the relationship between [NII]/H$\alpha$ and H$\alpha$ luminosity discovered by Crawford et al.~(1999), the observed uniformity of the line ratios we find in our sample may be due to the fact that, at high H$\alpha$ luminosity, all parts of the CCG are saturated at the lowest [NII]/H$\alpha$ values characteristic of massive star formation. In less H$\alpha$-luminous CCGs, we would thus predict more variations in the line ratio, with [NII]/H$\alpha$ increasing away from the less-abundant star-forming clumps. 

The possibility that galaxy interactions may trigger star-formation in these molecular gas-rich environments highlights the similarities between these CCGs and luminous infrared galaxies, as discussed by Edge~(2001). The {\em Spitzer} observations of Egami et al.~(2006) further strengthen the connection, revealing large thermal dust luminosities $L_{\rm{IR}} > 10^{11}$\Lsun~in two of the most H$\alpha$-luminous CCGs (Zw3146 and A1835).

\section{CONCLUSIONS}
The results presented in this paper have shed new light on the kinematic and morphological
complexity of the emission line gas in the cores of these H$\alpha$-luminous clusters. Our principal 
findings are that: $\bullet$ In many cases the H$\alpha$ emission appears to have been disturbed by interaction with a small companion galaxy 
to the central galaxy. $\bullet$ The CO and H$\alpha$ emission share the same kinematics and by implication trace each other closely. 
$\bullet$ The emission line ratios of the gas are uniform, showing no strong variations in [NII]/H$\alpha$, [SII]/H$\alpha$ or [OI]/H$\alpha$, as
a function of position or H$\alpha$ surface brightness, suggesting that its excitation state is independent of the processes which
disturb it on the observed kiloparsec scales.

Our interpretation is that the H$\alpha$ and CO emission are produced in a population of molecular clouds warmed by a starburst, which has itself 
been triggered by the passage of a secondary galaxy through the gas-rich cluster core. In these most luminous H$\alpha$ CCGs, this galaxy-wide
starburst leads to the saturation of the optical emission line ratios at a uniform level characteristic of massive star formation. In the absence 
of such triggering, the cooled gas reservoir will be in a more quiescent state with a much lower level of H$\alpha$ and CO emission; optical emission line ratios will be governed by other processes, e.g. excitation through cloud--cloud collisions (as described above) or mixing layers (Crawford \& Fabian~1992), to mention just two possibilities. Starting from this low H$\alpha$ luminosity quiescent state, adding some star-forming clumps will
increase the overall H$\alpha$ and CO luminosity and give rise to spatial variations in the optical line ratios. At the highest H$\alpha$ luminosities, star-formation will be widespread, with little spatial variation in optical emission line ratios.

To test this interpretation, future IFU observations should be carried out on CCGs covering the [OII]$\lambda 3727$ and 
[OIII]+H$\beta$ emission lines and the sub-4000\AA~continuum in order to probe current star formation, and with 
near-infrared IFUs to study the H$_{\rm{2}}$ and Pa$\alpha$ emission. Observations should also be performed on lower H$\alpha$ luminosity
CCGs in order to examine whether spatial variations in optical emission line ratios are indeed larger, due to less widespread star formation. 
On a longer timescale, mm-interferometers such as ALMA will reveal the spatial distribution of the CO emission (and dust) 
for a full comparison with the H$\alpha$. Our findings also highlight the need for numerical modelling of these dense 
CO reservoirs, with and without external perturbations.

\section*{ACKNOWLEDGEMENTS}
RJW and AMS are supported by PPARC. ACE acknowledges The Royal Society for 10 years of support during which much of the
groundwork for the present work was laid. The data published in this paper have been reduced using VIPGI designed by the 
VIRMOS Consortium and developed by INAF Milano, in connection with which we thank Bianca Garilli and Marco Scodeggio for 
their hospitality and assistance. We would also like to thank WHT staff for the OASIS IFU service observations, in particular 
Chris Benn and Danny Lennon, and Richard McDermid for data reduction advice. The William Herschel Telescope is operated on 
the island of La Palma by the Isaac Newton Group in the Spanish Observatorio del Roque de los Muchachos of the Instituto de 
Astrofisica de Canarias. We are grateful to Andy Fabian for the {\em Chandra} image of A1664.

{}

\end{document}